\documentclass[runningheads]{svmult}

\usepackage{graphicx}
\usepackage{amssymb}
\usepackage{amsmath}

\begin{document}

\title*{Asymptotic solutions of the nonlinear Boltzmann equation
for dissipative systems}
\titlerunning{Asymptotic velocity distributions for dissipative systems}
\author{M. H. Ernst\inst{1} \and R. Brito\inst{2}}
\institute{Instituut voor Theoretische Fysica, Universiteit Utrecht,
Postbus 80.195, 3508 TD Utrecht, The Netherlands. Email: {\tt
ernst@phys.uu.nl}  \and Depto de F\'{\i}sica Aplicada I, Universidad
Complutense, 28040 Madrid, Spain. Email: {\tt
brito@seneca.fis.ucm.es}\\}

\maketitle
\begin{abstract}
Analytic solutions $F(v,t)$ of the nonlinear Boltzmann equation in
$d$-dimensions are studied for a new class of dissipative models,
called inelastic repulsive scatterers, interacting through
pseudo-power law repulsions, characterized by a strength parameter
$\nu$, and embedding inelastic hard spheres ($\nu=1$) and inelastic
Maxwell models ($\nu=0$). The systems are either freely cooling
without energy input or driven by thermostats, e.g. white noise, and
approach stable nonequilibrium steady states, or marginally stable
homogeneous cooling states, where the data, $v^d_0(t) F(v,t)$ plotted
versus $c=v/v_0(t)$, collapse on a scaling or similarity solution
$f(c)$, where $v_0(t)$ is the r.m.s. velocity. The dissipative
interactions generate overpopulated high energy tails, described
generically  by stretched Gaussians, $f(c) \sim \exp[-\beta c^b]$ with
$0<b<2$, where $b=\nu$ with $\nu>0$ in free cooling , and
$b=1+\frac{1}{2} \nu$ with $\nu \geq 0$ when driven by white noise.
Power law tails, $f(c) \sim 1/c^{a+d}$, are only found in marginal
cases, where the exponent $a$ is the root of a transcendental
equation. The stability threshold depend on the type of thermostat,
and is  for the case of free cooling located at $\nu=0$.\\ Moreover we
analyze an inelastic BGK-type kinetic equation with an energy
dependent collision frequency coupled to a thermostat, that captures
all qualitative properties of the velocity distribution function in
Maxwell models, as predicted by the full nonlinear Boltzmann equation,
but fails for harder interactions with $\nu>0$.
\end{abstract}

\section{Introduction}

Classic kinetic
theory \cite{Waldm58,Chapman,resibois,cercignani-book,Uhl+Fo63}
deals with elastic particles with energy
conserving dynamics. The system is described by the single particle
distribution function, whose time evolution is governed by the
nonlinear Boltzmann equation. The asymptotic states of such systems
follow the universal laws of thermodynamics, and the distribution
function is the Maxwell Boltzmann distribution. This scenario does not
apply to dissipative systems, where energy is lost in inelastic
interactions.

In elastic systems the approach to asymptotic states is characterized
by a {\em kinetic} stage  of rapid relaxation in velocity space to a
locally homogeneous equilibrium state, followed by a {\em
hydrodynamic} stage of slow approach to a globally homogeneous
equilibrium state. The time scale in the kinetic stage is the mean
free time $t_{{mf}}$ between collisions. In the kinetic theory of
inelastic systems
\cite{campbell,jenkins,savage,haff,goldhirsch,goldstein-shapiro,goldhirsch-zanetti}
the type of decay depends on the energy supply to the dissipative
system. Without energy supply there is first a kinetic stage of rapid
relaxation on the time scale $t_{{mf}}$ to a locally homogeneous
adiabatic state, the {\em homogeneous cooling} state, described by
{\em scaling} or {\em similarity} solutions with a slowly changing
parameter, at least for weakly inelastic systems. With energy supply
the evolution is more similar to the elastic case with, however,
equilibrium states replaced by non-equilibrium steady states. The
velocity distributions in these adiabatic or steady states are very
different from a Maxwell Boltzmann distribution. The subsequent stage
of evolution involves transport phenomena and complex hydrodynamic
phenomena of clustering and pattern formation
\cite{goldhirsch-zanetti,Nagel}.

The interest in granular matter in general has strongly stimulated new
developments in the kinetic theory of granular fluids and gases, which
show surprising new physics. A granular fluid is a collection of small
or large macroscopic particles, with short range repulsive hard core
interactions, in which energy is lost in inelastic collisions, and the
system cools when not driven. When rapidly driven, gravity can be
neglected. The dynamics is based on binary collisions and ballistic
motion between collisions, which {\it conserve total momentum}. So
these systems can be considered to be a granular {\it fluid or gas}.
The prototypical model for these so-called rapid granular flows is a
fluid or gas of perfectly smooth mono-disperse inelastic hard spheres,
and its non-equilibrium behavior can be described by the nonlinear
Boltzmann equation
\cite{campbell,jenkins,savage,haff,goldhirsch,goldstein-shapiro,goldhirsch-zanetti}.
The inelastic collisions are modeled   by a {\em coefficient of
restitution} $\alpha \: \left(0<\alpha<1\right)$, where
$\left(1-\alpha^2\right)$ measures the degree of inelasticity.

This review focuses on the first stage of evolution, and studies the
velocity distribution $F\left(v,t\right)$ in spatially homogeneous
states of inelastic systems. For that reason most of the citations,
given in this article, only refer to kinetic theory studies of
$F\left(v,t\right)$, and not to studies of transport properties. The
revival
\cite{WM96,brey-dufty-santos,esipov,TvN+ME-granmat,BBRTvW01,piasecki,brey,MS00,brito+huthman,ignacio-II,BN-PK-00,Bobyl-00,CCG-00}
in kinetic theory of inelastic systems has been strongly stimulated by
the increasing sophistication of experimental techniques
\cite{SciAm-01,GM-exp}, which make direct measurements of velocity
distributions feasible in non-equilibrium steady states.  In this
review we also include inelastic generalizations \cite{ETB} of the
classical repulsive power law interactions
\cite{Chapman,resibois,cercignani-book}, which embed both the
inelastic hard spheres $\left(\nu=1\right)$, as well as the recently
much studied
\cite{Rome1,Rome2,BN+PK-6-11,BN+PK-02,ME+RB-EPL,ME+RB-rapid,ME+RB-fest,AB+CC-proof,Nienhuis,Droz,AS+ME-condmat}
inelastic Maxwell models $\left(\nu=0\right)$ in a single class of
models, parametrized by an exponent $\nu$. This exponent characterizes
the dependence of the collision frequencies on the energy of impact at
collision.

In fact, the kinetic theory for such models is of interest in its own
right, as the majority of inter-particle interactions in  macroscopic
systems involve some effects of inelasticity. Our goal is to expose
the generic  and universal features of the velocity  distributions in
dissipative fluids, and to compare them with conservative fluids to
highlight the differences.

A classical problem in kinetic theory is the possibility of
overpopulated high energy tails in velocity distributions
\cite{bobylev-BKW,krook+wu}, as many physical and chemical processes
only occur above a certain energy threshold. Consequently, this old
problem has received a new stimulus through a large amount of recent
theoretical and experimental studies on tail distributions in many
particle systems with inelastic interactions. From the point of view
of kinetic theory the intriguing question is, what is the generic
feature causing overpopulated tails, possibly even power law tails, in
systems of {\em inelastic} particles, how does the overpopulation
depend on the scattering cross sections, and on the different forms of
energy input. The generic feature is the mechanism for overpopulation,
and not the specific shape of the tails.

Finally, from the point of view of nonequilibrium steady states, the
structure of velocity distributions in elastic and dissipative
systems, including the high energy tail, is a subject of continuing
research, as the universality of the Gibbs' state of thermal
equilibrium is lacking outside thermal equilibrium, and a possible
classification of generic structures would  be of  great interest in
many fields of non-equilibrium statistical mechanics.

The {\em plan of the paper} is as follows: in Section 2 we discuss a
simple inelastic BGK- or single-relaxation-time model \cite{BGK} to
illustrate the phenomenon of power law tails. The exponent in the
algebraic tail depends qualitatively in the same manner on the degree
of inelasticity as in 2- and 3-dimensional Maxwell models. In Section
3 the nonlinear Boltzmann equation is constructed for inelastic
generalizations of the classical repulsive power law potentials, to
which we refer as Inelastic Repulsive Scatterers or IRS-models. This
is done for freely cooling as well as for systems driven by
thermostats or heat sources. Section 4 gives a systematic analysis for
the energy balance equation, it derives the nonlinear integral
equation for the scaling or similarity solution, denoted by
$f\left(c\right)$, and presents an asymptotic analysis of the high
energy tails in the form of stretched Gaussians, $f\left(c\right) \sim
\exp\left[-\beta c^b\right]$ where $b<2$. The method used can only be
applied to IRS-models with $\nu >0$, where the exponent
$b=b\left(\nu\right)$ is found as a simple function of $\nu$.  The
case of freely cooling Maxwell models ($\nu=0$) forms a borderline
case, discussed in Section 5. Here algebraic tails, $f\left(c\right)
\sim 1/c^{d+a}$, are found, where the exponent $a$ is determined by a
transcendental equation. It yields $a=a\left(\alpha\right)$ as a
function of the degree of inelasticity. We end with some perspectives
and conclusions.

\section{Inelastic BGK Model}
\subsection{Kinetic equations}

The goal of this section is to present in the nutshell of a simple
Bhatnagar-Gross-Krook (BGK) model \cite{BGK} a preview of many of the
qualitative features of velocity relaxation in homogeneous systems.

A crude scenario for the relaxation without energy input suggests that
the system will cool down due to inelastic collisions, and the
velocity distribution $F\left(v,t\right)$ will approach a Dirac delta
function $\delta \left(\vec{v}\right)$ as $t\to\infty$, while the
width  or r.m.s. velocity $v_0\left(t\right)$ of this distribution,
defined as $\langle v^2 \rangle = \frac{1}{2} d v_0^2 $, is shrinking.
With a constant supply of energy, the system can reach a
non-equilibrium steady state (NESS).

To model this evolution we use a simple BGK-type kinetic equation,
introduced in 1996 by Brey et al. \cite{BMD},
\begin{equation} \label{a1}
\partial_t F\left(v,t\right) -D\nabla_{\vec{v}}^2
F\left(v,t\right)= -\omega_\nu\left(t\right)
\left[F\left(v,t\right)-F_0\left(v,t\right)\right].
\end{equation}
We have added a heating term, $
-D\nabla_{\vec{v}}^2 F\left(v,t\right)$  to the usual BGK equation which
represents the heating by a white noise of strength $D$
\footnote{For a systematic discussion of driven systems, see
Section 3.3}. Here the mean collision frequency,
$\omega_\nu\left(t\right)= 1/t_{{mf}}$, is a function of the r.m.s.
velocity $v_0\left(t\right)$, chosen as $\omega_\nu=v_0^\nu$, in
preparation of section 3.1. The kinetic equation describes the
relaxation of $F\left(v,t\right)$ with a time-dependent rate
$\omega_\nu\left(t\right)$ towards a Maxwellian with a width
proportional to $v_0\left(t\right)$, defined by
\begin{equation} \label{a2}
F_0\left(v,t\right)= \left(\sqrt\pi \alpha v_0\right)^{-d}
\exp\left[-\left(v / \alpha v_0\right)
^2\right] \equiv \left(\alpha v_0\right)^{-d} \phi\left(c/\alpha \right),
\end{equation}
where $c=v/v_0$.  The constant $\alpha$ ($0<\alpha<1$) is related to
the inelasticity, $\gamma =
\textstyle{\frac{1}{2}}\left(1-\alpha^2\right)$, of the model, and the
totally inelastic limit $(\alpha \to 0)$ is ill-defined in this model,
as the mean energy is divergent for $\alpha =0$.

\subsection{Free Cooling ($D=0$)}
The cooling law of the mean square velocity $\langle v^2\rangle =
\textstyle{\frac{1}{2}}  dv_0^2$, or the granular temperature $T=v_0^ 2$, is
obtained by applying $\int d\vec{v} v^2\left(\dots\right)$ to \eqref{a1} with
$\omega_\nu=v_0^\nu$. The result is $\partial_t v_0 = -\gamma
v_0^{\nu+1}$, yielding
\begin{equation}\label{vo-nu}
v_0\left(t\right)= v_0\left(0\right) /\left[ 1+\nu\gamma t v_0^\nu(0)
 \right] ^{1/\nu}.
\end{equation}
The result is a homogeneous cooling law, $T\sim t^{-2/\nu}$, which
agrees with Haff's law \cite{haff} for $\nu=1$, corresponding to
inelastic hard spheres. Note that for negative $\nu$ the homogeneous
cooling law takes the form,
\begin{equation}\label{min-nu}
v_0\left(t\right)= v_0\left(0\right) \left[ 1-|\nu|\gamma t
/v_0^{|\nu|}(0) \right]^{1/|\nu|},
\end{equation}
i.e. for $t > t_s \equiv v_0^{|\nu|}/ |\nu|\gamma$ the r.m.s. velocity
and the mean kinetic energy become negative, which is {\it
unphysical}, and so are the BGK-models with $\nu<0$.

As indicated in the introduction, the long time behavior of
$F\left(v,t\right) $ in free cooling is determined by a scaling or
similarity solution of the form $F\left(v,t\right)= v_0^{-d}
\left(t\right) f\left(v/v_0\left(t\right)\right)$. We insert this
ansatz in \eqref{a1}, eliminate $\partial_t v_0= -\gamma v_0^{\nu+1}$,
and obtain the scaling equation,
\begin{equation}\label{a4}
c\frac{d}{dc} f +\left(d+a\right) f =\frac{a}{\alpha^d}
\phi\left(\frac{c}{\alpha}\right),
\end{equation}
where $a$ is defined as,
\begin{equation}\label{a5}
a=1/\gamma =2/\left(1-\alpha^2\right) .
\end{equation}
We also note that the scaling equation is independent of $\nu$. The
exact solution of this equation is:
\begin{equation}\label{a6}
f\left(c\right)= \frac{A}{c^{d+a}} + \frac{a}{\alpha^d \pi^{d/2}}
\left(\frac{1}{c^{d+a}}\right) \int_0^c du\, u^{d+a-1} \exp\left[
-u^2/\alpha^2\right].
\end{equation}
The integration constant $A$ is fixed by the normalizations,
\begin{equation}\label{a7}
\int d{\vec{c}} \{1,c^2\} f\left(c\right)=\{1,\textstyle{\frac{1}{2}} d\},
\end{equation}
where $d{\vec{c}} = \Omega_d c^{d-1} dc$ with $\Omega_d = 2 \pi^{d/2} /
\Gamma\left(d/2\right)$ being the surface area of a $d-$dimensional hyper-sphere.
The normalization integral converges only near $c\simeq 0$ if $A=0$.
Then the solution \eqref{a6} with $A=0$ is identical
 to the scaling form, obtained in \cite{BMD} for
$\nu=1$.

For velocities far above thermal, i.e. $c\gg 1$, the solution has a
power law tail,
\begin{equation}\label{a8}
f\left(c\right)\sim  \frac{a \alpha^a}{\pi^{d/2}}
 \left(\frac{1}{c^{d+a}}\right)
\int_0^\infty du\, u^{d+a-1} e^{-u^2}= \frac{a
\alpha^a\Gamma\left( \frac{d+a}{2}\right)}{2\pi^{d/2}}
 \left(\frac{1}{c^{d+a}}\right),
\end{equation}
with exponent $a=1/\gamma=  2/\left(1-\alpha^2\right)$, such that
$\langle {c^2} \rangle$ is bounded for $\alpha >0$. The exact solution
\eqref{a6}, including its high energy tail, is independent of the
exponent $\nu$, that determines the energy dependence of the mean
collision frequency $\omega_\nu = v_0^\nu $ in the BGK model.

As we shall see in Section 5, a similar heavily overpopulated tail,
$f\left(c\right)\sim 1/c^{d+a}$ with $d>1$, will also be found in
freely cooling Maxwell model with $\omega_0=1\ \left(\nu=0\right)$.
There the exponent $a\left(\alpha\right)$ takes in the elastic limit
($\alpha \to 1$) the form $a \simeq 1/\gamma_0 = 4d
/\left(1-\alpha^2\right)$. However, in the {\it general class} of
IRS-models with collision frequency $\omega_\nu \sim v_0^\nu$ with
$\nu>0$ the tails are {\it not} given by {\it power laws}, but by {\it
stretched Gaussians}, $f\left(c\right) \sim\exp \left[-\beta
c^b\right]$ with $0< b=b\left(\nu\right)<2$. These models will be
introduced  and discussed in Section 3.2.

\subsection{NESS ($D\neq 0$)}

Next we extend the result of \cite{BMD} to an inelastic  BGK-model
with $\nu \geq 0$, driven by white noise. By applying $\int d\vec{v}\,
v^2$ to \eqref{a1} we obtain the temperature balance equation at {\em
stationarity},
\begin{equation}\label{a9}
\partial_t v_0^2 =4D -2\gamma_0 v_0^{\nu+2}=0,
\end{equation}
where the collisional dissipation, $2\gamma_0 v_0^{\nu+2}$, is
compensated by energy input from the external white noise. Also note
that for $\nu <-2$ the fixed point solution $v_0\left(\infty\right)$
in \eqref{a9} still exists, but it is unstable \cite{ETB}. If
$v_0\left(0\right) <v_0\left(\infty\right) $, then $v_0\left(t\right)$
vanishes as $t \to \infty$, and if $v_0\left(0\right) >
v_0\left(\infty\right) $ then $v_0\left(t\right)$ diverges
\footnote{Thanks are due to E. Trizac and A. Barrat for pointing out
to us that the energy balance equations in this article have stability
thresholds that are $\nu-$dependent.}.

To obtain the solution $F\left(v,\infty\right)$ of \eqref{a1} in the
NESS we rescale $F\left(v,\infty\right) = v_0^{-d}\left(\infty\right)
f\left(v/v_0\left(\infty\right)\right)$ to the standard width $\langle
c^2\rangle =\textstyle{\frac{1}{2}} d$, substitute the rescaled form
in the kinetic equation, and eliminate $D$, using the stationarity
condition \eqref{a9} as well as \eqref{a5}. This yields the rescaled
equation in {\em universal} form,
\begin{equation}\label{a10}
\frac{1}{c^{d-1}} \frac{d}{dc} c^{d-1} \frac{d}{dc}
f\left(c\right) -2a f\left(c\right)
= -\frac{2a}{\alpha^d} \phi\left( \frac{c}{\alpha}\right),
\end{equation}
where the normalizations \eqref{a7} are imposed. The O.D.E. shows that
$f\left(c\right)$ is independent of the noise strength, $D$, and does
not contain any dependence on the initial data. This equation can be
solved exactly, and more details will be published in \cite{ETB}.
However, for the purpose of this section, we only want to extract from
the differential equation \eqref{a10} the asymptotic form of
$f\left(c\right)$. In that case we may neglect in \eqref{a10} the
inhomogeneity $\phi\left(c/\alpha\right) \sim
\exp\left[-\left(c/\alpha\right)^2\right]$ for $c\gg \alpha$, and find
the asymptotic solution for the BGK-model driven by white noise, i.e.
\begin{equation}
\label{a11}
  \begin{split}
& f\left(c\right)  \sim \exp\left[-\beta c^b\right] \\
& b=1;\ \beta=\sqrt{2a}= 2/\sqrt{1-\alpha^2} .
  \end{split}
\end{equation}
The constant $\beta$ is independent of the parameter $\nu$. The
exponentially decaying high energy tail is also a 'stretched'
Gaussian, which is overpopulated when compared to a Maxwellian, but
the overpopulation is much less heavy than is the freely cooling case
\eqref{a8} with an algebraic tail.

As we shall see in Section 4.3, a similar exponential high energy tail
will be found in the white noise driven Maxwell model ($\nu=0;\ b=1;\
\beta=\sqrt{8/\left(1-\alpha^2\right)}$), but not in the general class
of IRS-models, where the stretching exponent $b$ takes a value in the
interval $0< b=\overline{b}\left(\nu\right) \leq 2$.

\section{Basics of inelastic scattering models}
\subsection{Boltzmann equation as a stochastic process}
The nonlinear Boltzmann equation  for dissipative interactions in the
{\em homogeneous cooling state} can be put in a broader perspective,
that covers both elastic and inelastic collisions, as well as
interactions where the scattering of particles is described either by
conservative (deterministic) forces, or by stochastic ones. To do so
it is convenient to interpret the Boltzmann equation as a stochastic
process, similar to the presentations in the classical articles of
Waldmann\cite{Waldm58}, and Uhlenbeck and Ford \cite{Uhl+Fo63}, or in
Ulam's stochastic model \cite{Ulam} showing the basics of the approach
of a one-dimensional gas of elastic particles towards a Maxwellian
distribution.

Consider a spatially homogeneous fluid of elastic or inelastic
particles in $d-$dimensions, specified by their velocities
($\vec{v},{\vec{w}}, \cdots$), and interacting through binary
collisions, $\left(\vec{v},{\vec{w}}\right) \to
\left(\vec{v}^\prime,{\vec{w}}^\prime\right)$,  that are described by
transition probabilities. To describe fluids out of equilibrium the
total momentum $\vec{G}
=\textstyle{\frac{1}{2}}\left(\vec{v}+{\vec{w}}\right)$ needs to be
conserved in a binary collision. The outgoing or post-collision
velocities $\left(\vec{v}^\prime,{\vec{w}}^\prime\right)$ can be
parametrized in terms of the incoming velocities
$\left(\vec{v},{\vec{w}}\right)$, and an impact (unit) vector
$\vec{n}$, that is chosen  on the surface of a unit sphere with a
certain probability, proportional to the collision frequency
$\textsc{a} \left(\vec{g}, \vec{n}\right) = \textsc{a} \left({\rm g}
,\hat{\vec{g}} \cdot \vec{n}\right)$, that in general depends on the
relative speed $g =\left|\vec{v} - {\vec{w}}\right|$ of the colliding
particles, and the angle between $\vec{g}$ and $\vec{n}$, where
$\hat{\vec{a}}$ denotes a unit vector.

In this article we  consider the simplest case of modeling the
inelastic collisions through a velocity-independent coefficient of
restitution $\alpha \:\left(0<\alpha<1\right)$, where  the component
${{\rm g}_\parallel}\equiv \vec{g} \cdot \vec{n}$ is not only
reflected, as in elastic collisions, but also reduced in size by a
factor $\alpha$, i.e.
\begin{equation} \label{diss-coll}
{{\rm g}_\parallel}^\prime = -\alpha {{\rm g}_\parallel}.
\end{equation}
The components $\vec{g}_\bot = \vec{g} - {{\rm g}_\parallel} \vec{n}$,
orthogonal to $\vec{n}$, remain unchanged. More explicitly, with the
help of momentum conservation we find for the post-collision
velocities resulting from the direct collisions
$\left(\vec{v},{\vec{w}}\right)\to
\left(\vec{v}^\prime,{\vec{w}}^\prime\right)=\left(\vec{v}^*,
{\vec{w}}^*\right)$,
\begin{equation}
\label{v*}
  \begin{split}
\vec{v}^\ast =& \vec{v} -\textstyle{\frac{1}{2}}
\left(1+\alpha\right)\vec{g} \cdot \vec{n} \vec{n} \\
{\vec{w}}^\ast =& {\vec{w}} +\textstyle{\frac{1}{2}}
\left(1+\alpha\right)\vec{g} \cdot \vec{n} \vec{n}.
  \end{split}
\end{equation}
The corresponding energy loss in such a collision is,
\begin{equation} \label{E-loss}
\Delta E =\textstyle{\frac{1}{2}} \left(v^{*2}+w^{*2}-v^2 -
w^2\right)= -\textstyle{\frac{1}{4}}
\left(1-\alpha^2\right)\left(\vec{g} \cdot \vec{n}\right)^2,
\end{equation}
where $\left(1-\alpha^2\right)$ measures the degree of inelasticity. The
value $\alpha=1$ describes elastic collisions. In the special case
of hard spheres  the impact vector $\vec{n}$ is the unit vector along
the line of centers of the colliding spheres at contact, as
illustrated in Figure 1.

\begin{figure}[htbp]
$$\includegraphics[angle=270,width=.55 \columnwidth]{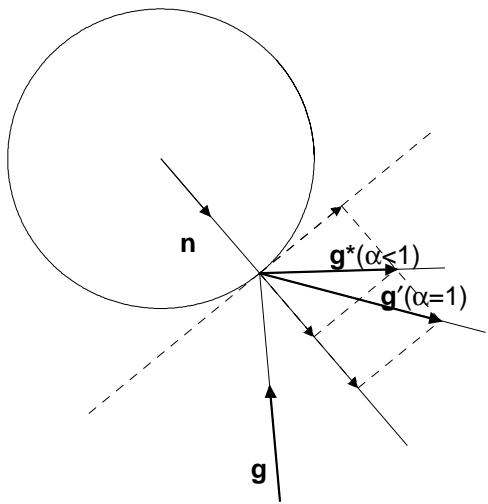}$$
\caption{Elastic ($\alpha=1$) and inelastic ($\alpha <1$)
scattering of hard spheres, where the parallel component ${{\rm g}_\parallel}$
is reflected as in \eqref{diss-coll}. In elastic collisions the
scattering angle $\chi = \cos^{-1}\left(\hat{\vec{g}}\cdot \hat{\vec{g}}^\prime\right)= \pi
-2 \phi $ with $\phi = \cos^{-1}\left(\hat{\vec{g}}\cdot \vec{n}\right)$ and in
inelastic ones $\chi = \cos^{-1}\left(\hat{\vec{g}}\cdot \hat{\vec{g}}^*\right)= \pi -
\phi - \phi^*$ with $\phi^* = \cos^{-1}\left(\hat{\vec{g}}^*\cdot \vec{n}\right)$.}
\end{figure}

The system above can be described by an isotropic velocity
distribution, $F\left(v,t\right)= F\left(\left|\vec{v}\right|,
t\right)$, as long as one only considers isotropic initial
distributions. Its time evolution is given by the Boltzmann equation,
\begin{multline}
  \label{BE-scatt}
\partial_t F\left(v,t\right)=  I\left(v|F\right)\equiv \int  d{\vec{w}}
d\vec{v}^\prime d{\vec{w}}^\prime\int d\vec{n} \left[W\left(\vec{v},
{\vec{w}}|\vec{v}^\prime,{\vec{w}}^\prime;\vec{n}\right) \right.\\
\left.\times F\left(v^\prime,t\right)F\left(w^\prime,t\right) -
W\left(\vec{v}^\prime,{\vec{w}}^\prime|\vec{v},{\vec{w}};\vec{n}
\right)F\left(v,t\right)F\left(w,t\right) \right].
\end{multline}
Here
$W\left(\vec{v}^\prime,{\vec{w}}^\prime|\vec{v},{\vec{w}};\vec{n}\right)$
is the transition probability per unit time that the incoming pair
state $\left(\vec{v},{\vec{w}}\right)$ at impact vector $\vec{n}$ is
scattered into the outgoing pair state
$\left(\vec{v}^\prime,{\vec{w}}^\prime\right)$. The loss term is the
sum over all  parameters of the direct collisions
$\left(\vec{v},{\vec{w}}\right) \to
\left(\vec{v}^\prime,{\vec{w}}^\prime\right)$  at fixed $\vec{v}$,
including a sum over impact vectors $\vec{n}$. Similarly the gain term
is the sum over all parameters of the restituting collisions,
$\left(\vec{v}^\prime,{\vec{w}}^\prime\right) \to \left(\vec{v}
,{\vec{w}} \right)$ at fixed $\vec{v}$. The transition probability for
the scattering event $\left(\vec{v},{\vec{w}}\right) \to
\left(\vec{v}^\prime,{\vec{w}}^\prime\right)$, obeying the inelastic
reflection law \eqref{diss-coll} and momentum conservation, is in
general proportional to the  collision frequency
$\textsc{a}\left(\vec{g},\vec{n}\right)$, and contains delta
functions, selecting the allowed collisions,
\begin{equation} \label{W-gen}
W\left(\vec{v}^\prime,{\vec{w}}^\prime|\vec{v},{\vec{w}}
;\vec{n}\right) =\textsc{a}\left(\vec{g},\vec{n}\right)
\delta^{\left(d\right)}\left(\vec{G}^\prime -
\vec{G}\right)\delta^{\left(d-1\right)}\left(\vec{g}^\prime_\bot
-\vec{g}_\bot\right) \delta\left({{\rm g}_\parallel}^\prime +
\alpha {{\rm g}_\parallel}\right).
\end{equation}
The faster the collision frequency increases at large impact
velocities, the more rapidly the high energy tail of
$F\left(v,t\right)$ relaxes relative to the bulk values of
$F\left(v,t\right)$ with $v$ in the thermal range.

For conservative interactions the total energy is conserved, as well
as total momentum and total number of particles. Moreover
\eqref{W-gen} shows that the transition probability is symmetric for
$\alpha=1$,
\begin{equation} \label{Dbalance}
W\left(\vec{v}^\prime,{\vec{w}}^\prime|\vec{v},{\vec{w}};
\vec{n}\right)= W\left(\vec{v},{\vec{w}}|\vec{v}^\prime,
{\vec{w}}^\prime;\vec{n}\right).
\end{equation}
This means that the transition probabilities for elastic binary
collisions satisfy the condition of {\it detailed balance}. Energy
conservation and the detailed balance relation in combination with the
$H-$theorem guarantee that the Maxwellian velocity distribution is
approached at large times.

Once the detailed balance relation is obeyed, it is trivial to prove
the $H-$theorem. Defining the $H-$function or entropy
$S\left(t\right) = - H\left(t\right)$
as,
\begin{equation} \label{H}
H\left(t\right) = \int d\vec{v} F\left(\vec{v},t\right)
\ln F\left(\vec{v},t\right),
\end{equation}
we obtain its decay rate by applying  $ \int d\vec{v}
\ln F\left(\vec{v},t\right)$ to
\eqref{BE-scatt}, then symmetrizing over $\vec{v} \leftrightarrow {\vec{w}}$ and
$\vec{v}^\prime \leftrightarrow {\vec{w}}^\prime$, and subsequently over
$\left(\vec{v},{\vec{w}}\right)
\leftrightarrow \left(\vec{v}^\prime,{\vec{w}}^\prime\right)$. The result is,
\begin{equation} \label{H-theorem}
\partial_t H\left(t\right) = \textstyle{\frac{1}{4}} \int d\vec{v}
d{\vec{w}} d\vec{v}^\prime d{\vec{w}}^\prime\int d\vec{n}
W \left[Y-X\right] \ln \left(X/Y\right) \leq 0,
\end{equation}
where $X= F\left(v\right)F\left(w\right)$ and $Y=
F\left(v^\prime\right)F\left(w^\prime\right)$
and the inequality follows
from $\left[Y-X\right] \ln \left(X/Y\right) \leq 0$.
The equality sign holds if and only if
$X=Y$. This implies that $H$ decreases monotonically and becomes
stationary only when $F$ approaches the Maxwellian.

 The reason for reviewing these 'obvious' properties in elastic
systems is that several of them, such as energy conservation, the
detailed balance relation,  the $H-$theorem, and the approach to a
stationary Maxwellian velocity distribution no longer hold in
dissipative systems $\left(\alpha<1\right)$, as we shall see.

\subsection{Boltzmann equation in standard form}
The standard form of the collision term $I\left(v|F\right)$  in the
Boltzmann equation for {\em elastic} interactions ($\alpha=1$)
contains the differential scattering cross-section $\sigma\left(
g,\chi\right)$. It can be calculated from the pair potential
$V\left(r\right)$ \cite{cercignani-book}, and depends on the relative
speed $g =\left|\vec{g}\right|$ and the scattering angle
$\chi=\pi-2\varphi$, where $\chi$ and $\varphi$ are defined in Figure
1. The collision frequency is then given by  $  \textsc{a}
\left(\vec{g},\vec{n}\right) = g \sigma\left( g,\chi\right)$. For
repulsive power law potentials $V\left(r\right) \sim r^{-s}$ the
collision frequency is $ \textsc{a}\left(\vec{g},\vec{n}\right) =
g^\nu\textsc{a}\left(\hat{\vec{g}}\cdot \vec{n}\right)$ with $\nu = 1
-2\left(d-1\right)/s$ in $d-$dimensions \cite{ME-PhysRep}. By
definition Maxwell molecules have a collision frequency, which is
independent of $g$, corresponding to $\nu=0$ or $s=
2\left(d-1\right)$. For hard spheres, $s \to \infty$ or $\nu=1$ with $
\textsc{a}\left(\vec{g},\vec{n}\right)= \left|\vec{g} \cdot \vec{n}
\right|\sigma^{d-1} =\left|{{\rm g}_\parallel}\right| \sigma^{d-1}$,
and $\sigma$ is the hard sphere diameter. Using these relations the
collision term in \eqref{BE-scatt} can be reduced to the classical
Boltzmann equation for elastic energy-conserving collisions. For {\em
inelastic} hard spheres the collision rate is again given by
$\textsc{a} \left(\vec{g},\vec{n}\right) = \left|{{\rm
g}_\parallel}\right|\sigma^{d-1}$. In general, a given positive
function $\textsc{a} \left(\vec{g},\vec{n}\right)$ defines a
stochastic scattering model, and in particular the choice, $\textsc{a}
\left(\vec{g},\vec{n}\right)=\textsc{a}\left(\hat{\vec{g}} \cdot
\vec{n}\right)$, defines inelastic Maxwell models
\cite{BN-PK-00,Bobyl-00,CCG-00}. As a simple realization of a
dissipative scattering model we consider the IRS-models with a
collision frequency $\textsc{a}\left(\vec{g},\vec{n}\right) =
\textsc{a}_0 \left|\vec{g} \cdot \vec{n}\right|^\nu $ with $\nu \geq
0$. This class includes inelastic hard spheres $\left(\nu=1\right)$
and inelastic Maxwell models $\left(\nu =0\right)$. In the remainder
of this article we restrict ourselves to this class of models.

To reduce the Boltzmann equation  to its standard form we use the
transition probabilities for dissipative interactions \eqref{W-gen}
with $\textsc{a}\left(\vec{g}, \vec{n}\right) =\textsc{a}_0
\left|{{\rm g}_\parallel}\right|^\nu$. Consider
first the loss term in \eqref{BE-scatt}, insert the relation
$d\vec{v}^\prime d{\vec{w}}^\prime = d\vec{g}^\prime_\bot
d{{\rm g}_\parallel}^\prime d\vec{G}^\prime$,
and carry out the integrations over the delta functions. The
result is,
\begin{equation} \label{loss}
\left(\partial_t F\left(v\right)\right)_{\mbox{loss}} =
- \int d{\vec{w}}\int d\vec{n}
\textsc{a}_0 \left|{{\rm g}_\parallel}\right|^\nu
F\left(w\right) F\left(v\right).
\end{equation}
The gain term involves the transition probability for the restituting
collisions $\left(\vec{v}^\prime,{\vec{w}}^\prime\right) \to
\left(\vec{v},{\vec{w}}\right)$, obtained from \eqref{W-gen} by
interchanging primed and unprimed velocities,
\begin{equation}
\label{gain}
  \begin{split}
\left(\partial_t F\left(v\right)\right)_{\mbox{gain}} =
&  \int d{\vec{w}} d\vec{g}^\prime_\bot
d{{\rm g}_\parallel}^\prime d\vec{G}^\prime \int d\vec{n}
\textsc{a}_0 \left|{{\rm g}_\parallel}^\prime\right|^\nu
\delta^{\left(d\right)}\left(\vec{G}-\vec{G}^\prime\right)\\
&\times \: \delta^{\left(d-1\right)}\left(\vec{g}_\bot -
\vec{g}^\prime_\bot\right)\frac{1}{\alpha} \delta\left(
{{\rm g}_\parallel}^\prime+ \frac{1}{\alpha}{{\rm g}_\parallel}\right)
F\left(w^\prime\right) F\left(v^\prime\right)\\
=& \int d{\vec{w}}\int d\vec{n}  \left(1/\alpha\right)\textsc{a}_0
\left|{{\rm g}_\parallel}/\alpha\right|^\nu
F\left(w^{**}\right) F\left(v^{**}\right).
  \end{split}
\end{equation}
In the second integral we have carried out the integrations over the
primed velocities, and used the following relations for the
restituting velocities,
\begin{equation}
\label{v**}
  \begin{split}
\vec{v}^\prime =&\vec{G}^\prime + \textstyle{\frac{1}{2}}
\vec{g}^\prime_\bot + \textstyle{\frac{1}{2}}
{{\rm g}_\parallel}^\prime \vec{n} = \vec{G} +
\textstyle{\frac{1}{2}}
\vec{g}_\bot - \textstyle{\frac{1}{2\alpha}}
{{\rm g}_\parallel} \vec{n} \\
=&\vec{v}-\textstyle{\frac{1}{2}} \left(1+
\textstyle{\frac{1}{\alpha}}\right)\vec{g} \cdot \vec{n} \vec{n}
\equiv \vec{v}^{**} \\
{\vec{w}}^\prime =& {\vec{w}} +
\textstyle{\frac{1}{2}}
\left(1+\textstyle{\frac{1}{\alpha}}\right)\vec{g} \cdot \vec{n}
\vec{n} \equiv {\vec{w}}^{**}.
  \end{split}
\end{equation}
In the first equality $\vec{v}^\prime$ has been expressed in center of
mass and relative velocities. In the second equality we have used the
inelastic collision law \eqref{diss-coll} and conservation of total
momentum, and the very last equality defines the restituting
velocities, $\left(\vec{v}^{**},{\vec{w}}^{**}\right)$. They are the
incoming velocities that result in the scattered velocities
$\left(\vec{v},{\vec{w}}\right)$, described by the inverse of the
transformation \eqref{v*}.

The space-homogeneous  Boltzmann equation  in its standard form is
then obtained by combining \eqref{loss} and \eqref{gain} with
$\textsc{a}\left(\vec{g},\vec{n}\right)= \left|\vec{g}
\cdot \vec{n}\right|^\nu$ to yield,
\begin{equation}
\label{BE-nu}
  \begin{split}
\partial_t F\left(v\right)=& I\left(v|F\right) \\
 I\left(v|F\right)=& \int_{\vec{n}}  \int d{\vec{w}}
 \left|\vec{g} \cdot \vec{n}\right|^\nu
\left[ \frac{1}{\alpha^{\nu+1}} F\left(v^{**}\right)
F\left(w^{**}\right) - F\left(v\right) F\left(w\right)
\right],
  \end{split}
\end{equation}
where $ \int_{\vec{n}} \left(\cdots\right) = \left(1/\Omega_d\right)
\int d\vec{n}  \left(\cdots\right)$ is an average over a
$d-$dimensional unit sphere, and we have absorbed constant factors in
the time scale. Here $\nu=1 $ corresponds to inelastic hard spheres
and $\nu=0$ to inelastic Maxwell models. Velocities and time have been
dimensionalized in terms of the width and the mean free time of the
initial distribution. Moreover, the Boltzmann equation obeys
conservation of particle number and total momentum, but the average
kinetic energy or granular temperature, $T\sim \left<v^2\right>$,
decreases in time on account of the dissipative collisions, i.e.
\begin{equation}\label{low-m}
\int d\vec{v} \left(1,\: \vec{v} ,\: v^2\right) F\left(v,t\right)
= \left(1,\:0,\:\textstyle{\frac{1}{2}} d v_0^2\left(t\right)\right),
\end{equation}
where $v_0\left(t\right)$ is the r.m.s. velocity. The inelastic
scattering models with collision frequency $\textsc{a}  \sim g^\nu
\:\left(\nu >0 \right)$, are the inelastic analogs of the
deterministic models with repulsive power law potentials,
$V\left(r\right) \sim r^{-s}$ with $2\left(d-1\right) < s < \infty$,
introduced at the start of this section. The inelastic case with
$\nu=2$ corresponds to an exactly solvable stochastic scattering model
with energy conservation, known as Very Hard Particle model
\cite{ME-PhysRep}.

The most basic and most frequently used model for dissipative systems
with short range hard core repulsion is the Enskog-Boltzmann equation
for inelastic hard spheres in $d-$dimensions \cite{campbell}, which
simplifies in the spatially homogeneous case to \eqref{BE-nu} with
$\nu=1$. Recently, inelastic Maxwell models have been studied
extensively. Ben-Naim and Krapivky \cite{BN-PK-00} introduced the
one-dimensional version of \eqref{BE-nu} with $\left(d=1,
\nu=0\right)$, and Bobylev et al \cite{Bobyl-00} have introduced  a
three-dimensional Maxwell model with $\textsc{a}
\left(\vec{g},\vec{n}\right) =\textsc{a}_0 \left|\hat{\vec{g}} \cdot
\vec{n}\right|$. The Maxwell models in \eqref{BE-nu} for general $d$
were first considered in \cite{BN+PK-6-11,ME+RB-EPL}.

\subsection{Cooling and driven systems}

\noindent{\em Homogeneous cooling and scaling:}\\
An inelastic fluid without energy input will cool down due to the
collisional dissipation in \eqref{E-loss}. In experimental studies of
granular fluids energy has to be supplied at a constant rate to keep
the system in a non-equilibrium steady state, while in analytic,
numerical and simulation work freely cooling systems can be studied
directly. Without energy input the velocity distribution $F\left(v,t\right)$ will
approach a Dirac delta function $\delta \left(\vec{v}\right)$ as $t \to \infty$, and
all moments approach zero, including the width $v_0\left(t\right)$.

However, an interesting structure is revealed when velocities, ${\vec{c}} =
\vec{v}/v_0\left(t\right)$, are measured in units of the instantaneous width
$v_0\left(t\right)$, and the long time limit is taken while keeping ${\vec{c}}$
constant, the so-called {\em scaling limit}. Monte Carlo simulations
\cite{Rome1} of the Boltzmann equation suggest that in this limit the
rescaled velocity distribution of the homogeneous cooling state can be
collapsed on a scaling form or similarity solution $f\left(c\right)$. These
observations seem to indicate that the long time behavior of $F\left(v,t\right)$
in freely cooling systems approaches a simple, and to some extent {\em
universal}, form $f\left(c\right)$, which is the same for different initial
distributions. Such {\em scaling} or {\em similarity} solutions  have
the structure,
\begin{equation} \label{f-scale}
 F\left(v,t\right) = \left(v_0\left(t\right)\right)^{-d}
f\left(v/v_0\left(t\right)\right),
\end{equation}
where ${\vec{c}} = \vec{v}/v_0\left(t\right)$ is the
scaling argument. Then $f\left(c\right)$
satisfies the normalizations,
\begin{equation} \label{norm-scale}
\int d{\vec{c}} f\left(c\right) =1; \quad \int d{\vec{c}}
c^2 f\left(c\right) =\textstyle{\frac{1}{2}} d,
\end{equation}
on account of \eqref{low-m}.  Substitution of the scaling ansatz
\eqref{f-scale} in the Boltzmann equation \eqref{BE-nu} then leads to a
{\em separation of variables}, i.e.
\begin{equation}
\label{separ-scale-eq}
  \begin{split}
I\left(c|f\right) =&  \gamma \left( df +c df/dc \right) =
\gamma \vec{\nabla}_{\vec{c}} \cdot \vec{c}f \\
\dot{v}_0 =& -\gamma v_0^{\nu+1},
  \end{split}
\end{equation}
where $\gamma$ is a separation constant.

The question is then: Can one determine physically acceptable
solutions? Here a short history. $\:$The first kinetic model with
dissipative interactions, that has been solved exactly for the case of
free cooling, {\em and} exhibits a heavily overpopulated high energy
tail, $f\left(c\right) \sim 1/c^{d+a}$, is the inelastic BGK model,
discussed in Section 2. Asymptotic solutions $\left(c\gg 1\right)$ of
the scaling equation for inelastic hard spheres
\cite{esipov,TvN+ME-granmat} have predicted the existence of
exponential {\em high energy tails}, $f\left(c\right) \sim
\exp\left[-\beta c\right]$. We talk about {\em tails}, {\em
over-populations} or stretched Gaussians in $f\left(c\right)$, when
the ratio of $f\left(c\right) $ and a Gaussian is an  increasing
function of $c$ at $c \gg 1$. The  predictions about high energy tails
were later confirmed in great detail by Monte Carlo simulations of the
long time solutions of the nonlinear Boltzmann equation for inelastic
hard spheres \cite{brey,MS00}, as well as  in further analytic work
\cite{BBRTvW01}.

Similarity solutions for freely cooling inelastic Maxwell models
\cite{BN-PK-00,Bobyl-00} where first studied  in terms of the scaling
variable, $\left(1-\alpha^2\right)t$, relating large times and small
inelasticities \cite{Bobyl-00}, but the   solutions obtained turned
out to be unphysical, i.e. {\em non-positive}. The first exact
positive similarity solution was found by Baldassarri et al.
\cite{Rome1} for the one-dimensional Maxwell model. It shows a
surprisingly strong high energy tail of algebraic type, $\sim 1/c^4$.
Using Monte Carlo simulations of the Boltzmann equation, these authors
also showed that rather general initial distributions  approach
towards this scaling form for long times. Algebraic tails,
$f\left(c\right) \sim 1/c^{d+a}$, for Maxwell models in higher
dimensions have been obtained analytically in
\cite{BN+PK-6-11,BN+PK-02,ME+RB-EPL,ME+RB-rapid,ME+RB-fest}, and the
actual approach in time towards such scaling forms has been studied
analytically in \cite{ME+RB-fest,AB+CC-proof} for general
classes of initial distributions.\\ \\

\noindent{\em Driving and steady states:}\\
Heating may be described by applying an external stochastic force to
the particles in the system, or by connecting the system to a
thermostat, which may be modeled  by a frictional force. For example,
the friction force $\gamma \vec{v}$ -- here with a negative friction
coefficient $\left(-\gamma \right)$ -- is called a Gaussian
thermostat. Complex fluids (e.g. granular) subject to such forces can
be described by the microscopic equations of motion for the particles,
$\dot{{\vec{r}}}_i =\vec{v}_i$, and $\dot{\vec{v}}_i = \vec{a}_i +
\tilde{\vec{\xi}}_i$ $\left(i=1,2, \cdots\right)$, where $\vec{a}_i$
and $\tilde{\vec{\xi}}_i$ are respectively the systematic and random
forces per unit mass. Here $\vec{a}_i$ contains frictional forces,
which may depend on velocity, and in the present case
$\tilde{\vec{\xi}}_i$ represents external noise (modeling energy
input), which is taken to be Gaussian white noise with zero mean, and
variance,
\begin{equation} \label{WN-strength}
\overline{ \tilde{\xi}_{i,\alpha}\left(t\right)
\tilde{\xi}_{j,\beta}\left(t^\prime\right)} = 2D
\delta_{ij}\delta_{\alpha\beta} \delta\left(t-t^\prime\right),
\end{equation}
where $\alpha,\beta$ denote Cartesian components, and $D$ is the noise
strength.  The Boltzmann equation for system driven in this manner
takes the form,
\begin{equation}
\label{BE-driven}
  \begin{split}
&\partial_t F\left(\vec{v}\right)+ {\cal F}F\left(\vec{v}\right) =
I\left(v|F\right) \\
&{\cal F} = \vec{\nabla}_{\vec{v}} \cdot \vec{a} -
D \nabla_{\vec{v}}^2.
  \end{split}
\end{equation}
A detailed derivation on how to include frictional and stochastic
forces  in kinetic equations can be found, for instance, in
\cite{ME-TvN-badhonnef,WM96,TvN+ME-granmat}.

When energy is supplied at a constant rate, driven systems can  reach
a NESS. Again there is the question about universality of these
asymptotic states as $t \to \infty$. Does the NESS depend on the
inelasticity, on the type of thermostat, and on the initial
distribution? The basic idea to show universality is always
essentially the same. Rescale the velocity distribution
$F\left(v,\infty\right)$ by measuring velocities in terms of their
typical size, i.e. the width $v_0\left(\infty\right)$,
\begin{equation} \label{f-NESS}
F\left(v,\infty\right) = \left(v_0\left(\infty\right)\right)^{-d}
f\left(v/v_0\left(\infty\right)\right),
\end{equation}
and analyze the scaling form $f\left(c\right)$. The scaling equation
for the NESS function $f\left(c\right)$ will be analyzed in Sections
4.2 and 4.3. Here we make some comments about what is known. The
scaling solutions for the NESS show again overpopulated tails in the
form of stretched Gaussians, albeit with a larger stretching exponent
$b$ than in free cooling. The scaling solution for inelastic hard
spheres driven by white noise was first analyzed in
\cite{TvN+ME-granmat}, predicting $f\left(c\right) \sim
\exp\left[-\beta c^b\right]$ with $b=3/2$. The experiments  of Rouyer
and Menon \cite{GM-exp,SciAm-01}, and of Aranson and Olafsen
\cite{GM-exp} seem to confirm the stretched Gaussian behavior with
$b=3/2$. Maxwell models, driven by white noise, were first studied in
\cite{BN-PK-00,CCG-00}, but did not give any predictions about
overpopulated high energy tails. The high energy tails for Maxwell
models, driven by white noise, were predicted in \cite{ME+RB-rapid} to
have exponential tails, $f\left(c\right) \sim \exp\left[-\beta
c\right]$. A number of more detailed analytical and numerical studies
about this driven Maxwell model in one dimension
\cite{Nienhuis,Droz,AS+ME-condmat,BN-PK-Springer} have appeared as
well. Moreover, as will be shown in Section 4.2, the integral equation
for the scaling form $f\left(c\right)$ in free cooling is identical to
the integral equation for the NESS distribution for a special
thermostat, provided $F\left(v,\infty\right)$ is also rescaled to the
same width as in \eqref{norm-scale}. How to determine the scaling form
$f\left(c\right)$ in free cooling and NESS will be described in
subsequent sections.

\subsection{Qualitative analysis}

In order to illustrate the rich behavior of the inelastic systems, we
start with the Boltzmann equation for the one-dimensional inelastic
Maxwell model \eqref{BE-nu} ($d=1,\nu=0$), where $F\left(v,t\right)$ satisfies,
\begin{equation} \label{BE-1d}
\partial_t F\left(v\right)- D \nabla_v^2 F\left(v\right) = I\left(v|F\right),
\end{equation}
and the collision term has the form,
\begin{equation}
\label{coll-1d}
  \begin{split}
I\left(v|F\right) =& \int dw \left[\frac{1}{\alpha}
F\left(v^{**}\right)F\left(w^{**}\right) -
F\left(v\right)F\left(w\right)
\right] \\
=& -F\left(v\right) + \frac{1}{p}\int  du
F\left(u\right)F\left(\frac{v-q u}{p}\right).
  \end{split}
\end{equation}
All velocity integrations extend over the interval
$\left(-\infty,+\infty\right)$. Here the outgoing velocities
$\left(v,w\right)$, and the incoming ones
$\left(v^{**},w^{**}\right)$
are according to \eqref{v**} related by,
\begin{equation} \label{dyn-1d}
v=q v^{**} +p w^{**}; \qquad w= p v^{**} +q w^{**}
\end{equation}
with $p= 1-q = \textstyle{\frac{1}{2}} \left(1+\alpha\right)$.
 By changing integration
variables  $w \to v^{**}=u$ with $dw =\left(\alpha/p\right)
 du$, and using the relation $w^{**} = \left(v-q u\right)/p$
one obtains the second equality in \eqref{coll-1d}. The normalization
of mass and  mean square velocity $\left<v^2\right> =
\textstyle{\frac{1}{2}} v_0^2$ are given by \eqref{low-m} for $d=1$.
The temperature balance equation is obtained from \eqref{BE-1d} as,
\begin{equation} \label{E-loss-1d}
\partial_t v_0^2 = 4D -2pq v_0^2,
\end{equation}
and describes the approach to the non-equilibrium steady state (NESS)
with a width $v_0^2\left(\infty\right) =2D/pq$, where the heating through
random forces, $\sim D$, is balanced by the collisional losses.

To understand the physical processes involved we first discuss in a
qualitative way the relevant limiting cases. Without the heating term
$\left(D=0\right)$, equation \eqref{BE-1d} reduces to the freely cooling
inelastic Maxwell model.

If one takes in addition the elastic limit $(\alpha \to 1$ or $q \to
0)$, the collision laws reduce in the {\em one-dimensional} case to
$v^{**} =w, w^{**}=v$, i.e. an exchange of particle labels; the
collision term vanishes identically; every $F\left(v,t\right) =
F\left(v\right)$ is a solution; there is no randomization or
relaxation of the velocity distribution through collisions, and the
model becomes trivial at the Boltzmann level of description, whereas
the distribution function in the presence of {\em infinitesimal}
dissipation $\left(\alpha \to 1\right)$ approaches a Maxwellian.

If we turn on the noise $\left(D \neq 0\right)$ at vanishing
dissipation $\left(q=0\right)$, the  collision term in \eqref{BE-1d}
vanishes, and the granular temperature  follows from \eqref{E-loss-1d}
as $v^2_0\left(t\right)=v^2_0\left(0\right)+4Dt$, and increases
linearly with time. With stochastic heating {\em and} dissipation
(even in infinitesimal amounts) the system reaches a NESS.

\subsection{Comments}
\noindent {\em Spatial dependence:}\\
When {\em spatial} dependence of the distribution function
$F\left({\vec{r}},\vec{v};t\right)$ is relevant, the collision term in
the Enskog-Boltzmann equation must be slightly modified, namely  the
angular integration $\int_{\vec{n}}$ over the full solid angle should
be replaced by $2 \int_{\vec{n}} \theta \left(-\vec{g} \cdot
\vec{n}\right)$, where the unit step function $\theta \left(x\right)$
restricts the $\vec{n}-$integration to the pre-collision hemisphere
with $\vec{g} \cdot \vec{n} <0$ (see Figure 1). In the spatially
uniform case both representations are identical because the
restituting velocities \eqref{v**} are even functions of $\vec{n}$.

\vspace{3mm}
\noindent {\em Origin of Maxwell models:}\\
In several papers \cite{Bobyl-00,BN+PK-6-11,ME+RB-fest,Nienhuis}
inelastic Maxwell models have been introduced as more or less {\em ad
hoc mathematical simplifications} of the nonlinear collision term for
inelastic hard spheres. This has been done by replacing the relative
velocity $g$ in the hard sphere collision frequency $\textsc{a}
\left(\vec{g}, \vec{n}\right) = \left|\vec{g} \cdot \vec{n}\right|$ by
its mean value, $\left<g\right> \textsc{a}\left(\vec{n}\right) \sim
v_0\left(t\right)\textsc{a}\left(\vec{n}\right) $, where
$v_0\left(t\right)$ is the root mean square velocity.

This procedure guarantees that the homogeneous cooling law for
inelastic Maxwell model constructed in this way, is identical to the
one for inelastic hard spheres, and given by Haff's law \cite{haff},
$T\left(t\right) \sim t^{-2}$. The construction of inelastic Maxwell
and IRS models, followed in the present article, is more in the spirit
of \cite{BN-PK-00}, i.e. by defining the collision term through
transition probabilities for the scattering process
$\left(\vec{v},{\vec{w}}\right) \leftrightarrow
\left(\vec{v}^\prime,{\vec{w}}^\prime\right)$ with the proper
constraints. \vspace{3mm}

 \noindent {\em Particles with pseudo-power law repulsion:}\\
A system of $N$ inelastic hard spheres is a model with microscopic
particles, characterized by positions and velocities, and interacting
via well-defined force laws, that can be studied by means of MD
simulations. However microscopic particles with dissipative
inter-particle forces, such as inelastic Maxwell molecules and
IRS-models, are stochastic models, only defined in $N-$particle
velocity space. Molecular Dynamics simulations can not be performed
for such models.

\vspace{3mm}
\noindent {\em Violation of $H-$theorem:} \\
As the detailed balance symmetry and the $H-$theorem are lacking in
dissipative interaction models, there is no guarantee that the entropy
$S\left(t\right)=-H\left(t\right)$  is {\em non-decreasing}. In fact
the solutions of the Boltzmann equation for such dissipative
interaction models approach \cite{ME+RB-fest} for long times to a
scaling form, defined in \eqref{f-scale}. By inserting such solutions
into \eqref{H} and anticipating the cooling law \eqref{vo-nu} in the
next section, one easily verifies that the entropy in the {\em
scaling} state keeps {\em decreasing} as $t$ becomes large, i.e.
\begin{equation} \label{H-chaos}
S \left(t\right) = -H\left(t\right)  \sim -\left(d/\nu\right) \ln t.
\end{equation}
This result is typical for pattern forming mechanisms in
configuration space, where spatial order or correlations are
building up, as well as in chaos theory, where the rate of
irreversible entropy production is negative on an attractor
\cite{Dorfman}. Moreover, there is no fundamental objection
against decreasing entropies in an open system in contact with a
reservoir, which is here the energy sink formed by the dissipative
collisions. The dynamics in $N-$particle velocity space
corresponds to a contracting flow $d\vec{v}^* d{\vec{w}}^* = \alpha d\vec{v}
d{\vec{w}} $ where $\alpha <1$, where the probability is contracting
onto an attractor. This is a well known phenomenon in chaos
theory.

\vspace{3mm}
\noindent{\em Modeling dissipation:}\\
There are many ways to model inelastic collisions that dissipate the
relative kinetic energy of colliding particles; e.g. by using Hertz'
contact law \cite{Brill-TP-badhonnef,chen-doolen-phys Lett},
visco-elastic media \cite{Hutter}, or  coefficients of normal and
tangential restitution \cite{campbell,goldstein-shapiro}. The
coefficients of restitution  may also depend on the relative speed of
the colliding particles \cite{Brill-TP-badhonnef}.

For the IRS-models, studied in this article, the scattering laws
(scattering angle, collisional energy loss) are independent of $\nu$,
i.e. they are the same for inelastic hard spheres, inelastic Maxwell
models and for general IRS-models. Only the collision frequency
$\textsc{a} ({\bf g, n}) $ of the IRS-models has the same energy
dependence as the collision frequency of elastic particles interacting
through repulsive power law potentials.

\section{Analysis of inelastic scattering models}
\subsection{Homogeneous cooling laws}
We start with the freely evolving case and compute the cooling rate,
defined through, $\partial_t v_0^2 =- \zeta_\nu \left(t\right) v_0^2$.
This can be done by applying $\left(\int d\vec{v} v^2\right)$ to the
Boltzmann equation \eqref{BE-nu}, changing integration variables
$\left(\vec{v},{\vec{w}}\right)\to
\left(\vec{v}^{**},{\vec{w}}^{**}\right)$ in the gain term, and using
the relations $d\vec{v} d{\vec{w}} = \alpha d\vec{v}^{**}
d{\vec{w}}^{**}$ together with the energy loss $\Delta E$  per
collision in \eqref{E-loss}. In general the above cooling equation is
a formal identity, and does not provide a closed equation for
$v_0(t)$. However, if $F(v,t)$ rapidly relaxes to a scaling form, then
the {\it subsequent} time evolution of $v_0^2 =(2/d) \langle {v^2}
\rangle$ is described by a closed equation for $v_0(t)$, as we will
see, i.e.
\begin{multline}
\label{m2-gen}
\partial_t { v_0^2}  =\textstyle{\frac{2}{d}} \int d\vec{v} v^2
I\left(v|F\right) = \textstyle{\frac{2}{d}} \int_{\vec{n}}
\int d\vec{v} d{\vec{w}}
\left|\vec{g}\cdot \vec{n}\right|^\nu \Delta E F\left(v,t\right)
F\left(w,t\right) \\ =
-\frac{1}{2d}\left(1-\alpha^2\right) v_0^{\nu+2}\int _{\vec{n}}
\int d{\vec{c}} d{\vec{c}}_1
\left|\left({\vec{c}}-{\vec{c}}_1\right)\cdot\vec{n}\right|^{\nu+2}
f\left(c\right) f\left(c_1\right),
\end{multline}
from which the cooling rate can be identified as,
 \begin{equation} \label{zeta-nu}
\zeta_\nu\left(t\right) \equiv - \partial_t v_0^2/v_0^2 = 2\gamma_0
\kappa_{\nu+2}\:v_0^{\nu}\left(t\right).
\end{equation}
Here the coefficient $\gamma_0= \frac{1}{4d}\left(1-\alpha^2\right)$
measures the inelasticity. Moreover, the constant $\kappa_\nu$  is
defined as,
\begin{equation} \label{kappa-nu}
\kappa_\nu = \int _{\vec{n}} \int d{\vec{c}} d{\vec{c}}_1
\left|\left({\vec{c}}-{\vec{c}}_1\right)\cdot\vec{n}\right|^{\nu}
f\left(c\right) f\left(c_1\right)= \beta_\nu \int d{\vec{c}}
d{\vec{c}}_1 |{\vec{c}}-{\vec{c}}_1|^\nu f\left(c\right)
f\left(c_1\right),
\end{equation}
and the $\left(d-1\right)-$dimensional angular integral $\beta_\nu $
is evaluated as,
\begin{equation} \label{beta-nu}
\beta_{\nu} =\int_{\vec{n}} \left|\hat{\vec{a}}\cdot\vec{n}\right|^{\nu} =
\frac{\int_0^{\pi/2} d\theta \left(\sin \theta\right)^{d-2}
\left(\cos\theta\right)^{\nu} } {\int_0^{\pi/2} d\theta
\left(\sin\theta\right)^{d-2}}
=\frac{\Gamma\left(\frac{\nu+1}{2}\right)
\Gamma\left(\frac{d}{2}\right)}
       {\Gamma\left(\frac{\nu+d}{2}\right)\Gamma\left(\frac{1}{2}\right) }.
\end{equation}
In one-dimensional systems $\beta_\nu =1$ for all $ \nu$, and in
higher dimensions for the $\nu$-values of interest here  $\beta_\nu $
is convergent ($\nu>0$). As a consequence of the scaling ansatz the
coefficient $\kappa_\nu$ is a time independent constant, that does
depend on the unknown scaling form. For Maxwell models
$\left(\nu=0\right)$ the cooling rate $\zeta_0$ can be calculated
explicitly from \eqref{beta-nu}, and the constant $\kappa_2$ is
$\kappa_2= \left(1/d\right)
\left<\left|{\vec{c}}-{\vec{c}}_1\right|^2\right>=1$ on account of
\eqref{norm-scale}. The result is,
\begin{equation} \label{zeta-0}
\zeta_0 \equiv - \partial_t v_0^2/v_0^2 = 2\gamma_0.
\end{equation}
The r.m.s. velocity decays then as $v_0\left(t\right)=v_0\left(0\right)
 \exp\left[-\gamma_0 t\right]$.

The most important time scale in kinetic theory is the mean free time
$t_{{mf}}$, which equals the inverse of the {\em mean collision rate}
$\omega_\nu\left(t\right)$, defined as the average of the collision
frequency $\textsc{a}\left(\vec{g},\vec{n}\right)=
\left|\left({\vec{c}}-{\vec{c}}_1\right) \cdot \vec{n}\right|^\nu$
over the velocities of the colliding pair in the scaling state, i.e.
\begin{equation} \label{collfreq-nu}
\omega_\nu \left(t\right) \equiv -\int d\vec{v}
I_{\mbox{loss}}\left(v|F\right)=   \kappa_\nu
v_0^{\nu}\left(t\right).
\end{equation}
Comparison of \eqref{zeta-nu} and \eqref{collfreq-nu} shows that both
frequencies are related as,
\begin{equation}\label{gamma-nu}
\zeta_\nu \left(t\right) = 2 \gamma_0 \frac{\kappa_{\nu+2}}{\kappa_\nu}
\omega_\nu\left(t\right)\equiv 2 \gamma_\nu\omega_\nu \left(t\right),
\end{equation}
where $\gamma_\nu \kappa_\nu= \gamma_0 \kappa_{\nu+2}$ are time
independent constants.  Using \eqref{collfreq-nu} and \eqref{gamma-nu}
the equation for the r.m.s. velocity becomes, $\partial_t v_0 =
-\gamma_\nu\kappa_\nu v_0^{\nu+1}$, yielding a solution, identical to
\eqref{vo-nu} with $\gamma$ replaced by $\gamma_\nu$. Consequently the
granular temperature at large times decays as $T=v_0^2  \sim
t^{-2/\nu}$. For $\nu=1$ one recovers the well known law of Haff
\cite{haff}, describing  the long time decay of the granular
temperature in the homogeneous cooling state of inelastic hard
spheres. A homogeneous cooling law similar to \eqref{vo-nu} with
$\nu=6/5$ has been derived in \cite{Brill-TP-badhonnef}, not on the
basis of the Boltzmann equation \eqref{BE-nu} with an energy dependent
collision rate $\propto g^{5/6}$, but by using the Boltzmann equation
for inelastic hard spheres $\left(\nu=1\right)$ with an energy
dependent coefficient of restitution $\alpha\left(g\right) = 1-
Ag^{2/5}$. For Maxwell models $\left( \nu \to 0\right)$ the r.m.s.
velocity in \eqref{vo-nu} reduces to $v_0\left(t\right) =
v_0\left(0\right) \exp\left[-\gamma_0 t\right]$, in agreement with
\eqref{zeta-0}.

The homogeneous cooling law for the general class of IRS-models
discussed here, can be cast into a universal form by changing to a new
time variable, the {\em collision counter}  or {\em internal} time
$\tau$ of a particle, which represents the total number of collisions
that a particle has suffered in the (external) time $t$. It is defined
through the mean collision frequency,
\begin{equation} \label{dtau}
d\tau = \omega_\nu\left(t\right)dt,
\end{equation}
where $\omega_\nu(t) \sim v_0^\nu(t)$. After inserting \eqref{vo-nu}
in this equation the differential equation can be solved to yield,
\begin{equation} \label{tau}
\nu \gamma_\nu \tau  = \ln \left[1 + \nu \gamma_\nu
\omega_\nu\left(0\right) t \right],
\end{equation}
valid for all IRS-models. Combination of \eqref{tau} and \eqref{vo-nu}
shows that the r.m.s. velocity follows the {\em universal} homogeneous
cooling law,
\begin{equation} \label{vo-univ}
 v_0\left(\tau\right) = v_0\left(0\right)
\exp\left[ -\gamma_\nu \tau\right],
\end{equation}
where $\gamma_\nu$ is according to \eqref{gamma-nu} and
\eqref{kappa-nu} proportional to the inelasticity $\gamma_0$, and
depends on $\nu$ through the collision integrals $\kappa_\nu$. We
further observe that the relations \eqref{tau} and \eqref{vo-univ}
also holds for the inelastic BGK-model in Section 2 with $\gamma_\nu $
replaced by $\gamma= \textstyle{\frac{1}{2}} \left(1-\alpha^2\right)$
in \eqref{a5}.

So far we have been dealing with freely cooling systems. Next we
address the {\em balance equation} for the granular temperature in
driven cases, where the external input of energy counterbalances the
collisional cooling, and may lead to a NESS. We proceed in the same
manner as for the free case, and apply $\left(\int d\vec{v} v^2\right)$ to the
Boltzmann equation in \eqref{BE-driven} with the result,
\begin{equation} \label{T-balance}
\partial_t \left< v^2\right> = -\zeta_\nu\left(t\right)
\left< v^2\right> +2 \left<\vec{v} \cdot \vec{a}\right>
+2dD,
\end{equation}
where the first term, $\int d\vec{v} v^2 I\left(v|F\right) = -
\zeta_\nu \left(t\right)\left< v^2 \right>$, is obtained from
\eqref{m2-gen}, \eqref{zeta-nu} and \eqref{low-m}. The next two terms
are obtained from the driving term in \eqref{BE-driven}, i.e. $\int
d\vec{v} v^2 {\cal F} F\left(v\right)$, by performing partial
integrations.

The most common ways of driving dissipative fluids
\cite{TvN+ME-granmat,MS00,BN-PK-00,CCG-00} is by Gaussian white noise
(WN) $\left(\vec{a}=0; D \neq 0\right)$, or by a Gaussian thermostat
(GT) $\left(\vec{a} = \gamma \vec{v}; D=0\right)$, yielding for the
balance equations,
\begin{equation}
\label{vo-driven}
\partial_t v_0^2\left(t\right) =
\begin{cases}
\left(2\gamma -\zeta_\nu\left(t\right)\right)
v_0^2\left(t\right) &  \text{(GT)} \\
4D- \zeta_\nu \left(t\right) v_0^2\left(t\right) & \text{(WN).}
\end{cases}
\end{equation}
Here the collisional loss, $-\zeta_\nu v_0^2$, is counterbalanced  by
the heat, $2\gamma v_0^2$, generated by the negative friction of the
Gaussian thermostat, or by the heat, $4D$, generated by randomly
kicking the particles.

As $t \to \infty$ the granular temperature will reach a NESS
 with $T\left(\infty\right) = v_0^2\left(\infty\right)$,
where the r.m.s. velocity  follows
from \eqref{vo-driven} using \eqref{zeta-nu},
\begin{equation}
\label{vo-NESS}
v_0\left(\infty\right) =
\begin{cases}
\left(\gamma/\gamma_0 \kappa_{\nu+2}\right)^{1/\nu} &  \text{(GT)} \\
 \left(2D/\gamma_0 \kappa_{\nu+2}\right)^{1/\left(\nu+2\right)}
& \text{(WN)}.
\end{cases}
\end{equation}
With reference to the discussion in Section 2.3 we note  that the
GT-fixed point solution $v_0\left(\infty\right)$ is {\it attracting}
for $\nu>0$, leading to a stable NESS, and {\it unstable} for $\nu<0$
with $v_0\left(t\right)$ vanishing at $t \to \infty$ if
$v_0\left(0\right) < v_0\left(\infty\right)$, and diverging if
$v_0\left(0\right) > v_0\left(\infty\right)$. The case $\nu=0$ is {\it
marginally stable}. For the WN-fixed point similar observations apply
with the stability threshold $\nu=0$ replaced by $\nu=-2$.

\subsection{Scaling and non-equilibrium steady states}

\noindent {\em Free cases:\\} To investigate the existence of scaling
solutions of the Boltzmann equation for freely cooling inelastic
systems, we substitute  the scaling ansatz \eqref{f-scale} into
\eqref{BE-nu}, to obtain an integral equation for the scaling form
$f\left(c\right)$. With the help of \eqref{dtau}, \eqref{vo-univ} and
\eqref{separ-scale-eq}
 the left hand side of \eqref{BE-nu} becomes,
 \begin{multline}
   \label{lhs-scale}
\mbox{l.h.s.} =- {\dot{v}_0}{ v_0^{-d-1}}
\left(\frac{d\tau}{dt}\right) \left\{ df\left(c\right)+c\frac{d}{dc}
f\left(c\right)\right\}\\
= \gamma_\nu \omega_\nu\left(\tau\right)  v_0^{-d}\left(\tau\right)
\nabla_{{\vec{c}}}\cdot {\vec{c}} f\left(c\right) =
\gamma_0 \kappa_{\nu+2}
\nabla_{{\vec{c}}}\cdot{\vec{c}} f\left(c\right)\:
v_0^{\nu-d}\left(\tau\right).
 \end{multline}
In the last equality we have used the relation $\gamma_\nu\kappa_\nu =
\gamma_0 \kappa_{\nu+2}$, implied by \eqref{gamma-nu}. The resulting
integral equation for $f\left(c\right)$ becomes,
\begin{equation}\label{eq-f-scale}
I \left(c|f\right) = \gamma_0 \kappa_{\nu+2} \nabla_{{\vec{c}}}
\cdot{\vec{c}} f = \frac{1}{d}
\varpi_2 \nabla_{{\vec{c}}}\cdot{\vec{c}} f .
\end{equation}
With the help of \eqref{m2-gen} and \eqref{zeta-nu} the second moment of
the collision term, $\varpi_2$, can be expressed as,
\begin{equation}\label{m2-I-scale}
\varpi_2\equiv -\int d{\vec{c}} c^2 I\left(c|f\right) =
d\gamma_0\kappa_{\nu+2}.
\end{equation}

\noindent {\em Driven cases:}\\
We consider equation \eqref{BE-driven} for the NESS distribution in its
rescaled form \eqref{f-NESS} with $v_0= v_0\left(\infty\right)$, driven by a
Gaussian thermostat (GT), $\{ \vec{a}=\gamma\vec{v} ; D=0\}$ or by white
noise (WN), $\{ \vec{a}=0 ; D\neq 0\}$. The scaling equations for
$f\left(c\right)$ take the form,
\begin{equation}
\label{eq-f-driven}
I\left(c|f\right) =
\begin{cases}
  \displaystyle \frac{\gamma}{v_0^\nu} \nabla_{{\vec{c}}}\cdot{\vec{c}} f =
\frac{1}{d} \varpi_2 \nabla_{{\vec{c}}}\cdot {\vec{c}} f & \text{(GT)} \\
\displaystyle -\frac{D}{v_0^{\nu+2}} \nabla_{{\vec{c}}}^2 f = -\frac{1}{2d}
\varpi_2 \nabla_{{\vec{c}}}^2 f & \text{(WN)}\,.
\end{cases}
\end{equation}
The first equality in (GT) and (WN) suggests that $f\left(c\right)$ depends
explicitly on $\gamma$ or $D$. This is however not the case. The
stationarity relation \eqref{vo-NESS} combined with \eqref{m2-I-scale}
shows in fact that the following expressions,
\begin{equation}\label{NESS-cond}
  \varpi_2 = d\gamma v_0^{-\nu} = 2d D v_0^{-\left(\nu+2\right)} =d
\gamma_0 \kappa_{\nu+2},
\end{equation}
are {\em independent} of $\gamma$ or $D$. So, we have used
\eqref{vo-NESS} to eliminate $\gamma$ and $D$, and to put it in the
universal form, containing $\varpi_2$.

\subsection{Comments}

\noindent {\em Equivalence free cooling - Gaussian thermostat:}\\
Comparison of integral equation \eqref{eq-f-scale} for free cooling and
\eqref{eq-f-driven} for the Gaussian thermostat shows that both
equations are identical, as first observed by Montanero and Santos
\cite{MS00}. This implies that the scaling form $f\left(c\right)$ in free cooling
(${\cal F}=0$) is identical to the NESS distribution $f\left(c\right)$ of the
same system, driven by a Gaussian thermostat, provided both forms are
rescaled to the same constant width $\int d{\vec{c}} c^2 f\left(c\right)=d/2$. This
also implies that the scaling form for the free case can be measured
by performing Monte Carlo simulations in a steady state
\cite{brey,MS00}. The same idea of systematically rescaling the
velocities has also been used in molecular dynamics simulations of a
freely cooling system of $N$ inelastic hard spheres \cite{Soto}.

\vspace{3mm} \noindent {\em Universality:}\\  The scaling equations
\eqref{eq-f-driven} for $f\left(c\right)$ in systems driven by a Gaussian
thermostat and by white noise, have a universal form, because
$\varpi_2$ is independent of the friction constant $\gamma$, the noise
strength $D$ and the width $v_0\left(\infty\right)$, which may depend on the
initial distribution. The scaling equations reduce for $\nu =1$ to
those for hard spheres \cite{TvN+ME-granmat}, and for $\nu =0$ to
those for Maxwell models \cite{ME+RB-rapid}.

\vspace{3mm} \noindent {\em Perturbative approach:}\\
The moment of the collision term $\varpi_2=d\gamma_0 \kappa_{\nu+2}$
in the scaling equations \eqref{eq-f-driven}, and the cooling rate
$\zeta_\nu=\gamma_0\kappa_{\nu+2} v_0^\nu$ in \eqref{vo-driven}
contain for all models the quantity $\kappa_\nu$, as given by
\eqref{zeta-nu}. It depends on the unknown function $f\left(c\right)$,
except for $\nu=0$ where $\kappa_2=1$ and $\varpi_2=d \gamma_0
=\textstyle{\frac{1}{4}} \left(1-\alpha^2\right) $. For the case of
inelastic hard spheres, a perturbative method has been developed in
\cite{goldstein-shapiro,TvN+ME-granmat} for small inelasticities to
solve both integral equations in \eqref{eq-f-driven} by expanding
$f\left(c\right)$ in a series of Sonine polynomials
$S_p\left(c^2\right)$, i.e.
\begin{equation}\label{f-sonine}
f\left(c\right)= \phi\left(c\right) \{ 1+\sum_{p=2}^\infty a_p
S_p\left(c^2\right)\},
\end{equation}
where $\phi\left(c\right)= \pi^{-d/2} \exp\left(-c ^2\right)$ is the
Maxwellian. Here $a_2$ is essentially the fourth cumulant which has
been calculated explicitly in \cite{TvN+ME-granmat}. For dimensions
$d>1$ it turns out to be proportional to the inelasticity
$\left(1-\alpha^2\right)$. However, the one-dimensional case is
exceptional because $a_2$ approaches a finite value for $\alpha \to 1$
\cite{BBRTvW01}. The same method has been successfully applied by
Cercignani et al. \cite{CCG-00} to an inelastic Maxwell model, and by
several authors \cite{brey,MS00,brito+huthman,Brill-TP-badhonnef} to
inelastic hard spheres and related problems. The method  can be
applied to the inelastic IRS-models as well.

The method focuses on the lower moments of $f\left(c\right)$, and the
polynomial approximation, cut off after $S_2\left(c^2\right)$, gives a
fair representation of $f\left(c\right)$ for velocities in the thermal
range, $c\leq 2$. The expansion \eqref{f-sonine} can then be used to
calculate $\kappa_\nu$ in \eqref{m2-I-scale} in the form
$\kappa_\nu=\kappa_\nu^0+\left(1-\alpha^2\right) \kappa_\nu^1+\cdots$.
For more details we refer to \cite{TvN+ME-granmat}. As an illustration
we calculate the lowest approximation $\kappa_\nu^0$ to $\kappa_\nu$
 in \eqref{zeta-nu} using $f\left(c\right)\simeq\phi\left(c\right)$. The result is,
\begin{eqnarray} \label{kappa-0}
  \kappa_\nu^0 &=&\int_{\vec{n}} \int d{\vec{c}} d{\vec{c}}_1
\left|g_\parallel\right|^\nu\phi\left(c\right)\phi\left(c_1\right)\nonumber
\\ &=&\frac{1}{\sqrt{2\pi}}
\int_{-\infty}^{\infty} dx \left|x\right|^\nu e^{-x^2 /2}=
\frac{2^{\nu/2}}{\sqrt{\pi}} \Gamma\left(\frac{\nu+1}{2}\right).
\end{eqnarray}
The lowest approximation to \eqref{NESS-cond} is then $\varpi_2^0 =
d\gamma_0\kappa_\nu^0$. The damping rate $\gamma_\nu$ in
\eqref{gamma-nu} of the r.m.s. velocity becomes then
$\gamma_\nu=\left(\nu+1\right)\gamma_0$, in agreement with the results
for $\nu=1$ in \cite{TvN+ME-granmat}.

\subsection{ High energy tails}

The polynomial expansion \eqref{f-sonine} describes $f\left(c\right)$
only in the thermal range,  but contains no meaningful information
about velocities $c$ in the asymptotic range $\left(c \gg 1\right)$.
However the high energy tails in the IRS-models with $\nu>0$ can be
determined by a procedure similar to the one used successfully for
inelastic hard spheres systems \cite{TvN+ME-granmat}, as well as in
inelastic Maxwell models  driven by white noise \cite{ME+RB-rapid}. To
do so we make the ansatz of stretched Gaussian  behavior for the high
energy tail, i.e. ${f}\left(c\right)\simeq B \exp\left[-\beta
c^b\right]$ with $0<b<2$ and $ \beta >0$, and determine $b$ and
$\beta$ by inserting this ansatz in the scaling equation
\eqref{eq-f-driven} and requiring self-consistency. The border line
case $b \to 2$ corresponds to Gaussian tails, and $b \to 0$ suggests
power law tails with negative exponents.

An estimate of the rescaled collision term $I\left(c|f\right)$ in
\eqref{BE-nu} is made in \cite{TvN+ME-granmat,ME+RB-rapid}. This
suggests that the loss term is asymptotically dominant over the gain
term as long as the exponent $b$ in $\exp\left[ -\beta c^b\right]$ is
restricted to $b >0$. This estimate also applies to the IRS-models as
long as the exponent $\nu >0$ and the inelasticity $\gamma_0$ is
non-vanishing. So the gain term is neglected. Moreover the loss term
can be simplified in the asymptotic velocity range. Its dominant
contribution comes from collisions with particles having velocities
$c_1$ that are typically in the thermal range $\left( c_1 ={\cal
O}\left(1\right)\right)$. Consequently the collision rate
$\left|\left({\vec{c}} -{\vec{c}}_1\right) \cdot \vec{n}\right|^\nu$
for asymptotic dynamics may be replaced by  $ c^\nu
\left|\hat{{\vec{c}}}\cdot \vec{n}\right|^\nu$, and the total
collision term simplifies to,
\begin{equation} \label{I-asym}
I\left(c|f\right)\sim I_{\mbox{loss}}\left(c|f\right) \sim  - c^\nu
\beta_ \nu f\left(c\right),
\end{equation}
with $\beta_\nu$ given by \eqref{beta-nu}. After these preparations we
insert \eqref{I-asym} and the stretched Gaussian form,
${f}\left(c\right) \sim \exp\left[-\beta c^b\right]$, into the
Boltzmann equation \eqref{eq-f-driven}, and follow the procedure
sketched above. This gives the following universal results (see item 2
below \eqref{NESS-cond}) for the asymptotic high energy tail in
$d-$dimensional inelastic $\nu-$models, $f\left(c\right) \sim
\exp\left[-\beta c^b\right]$ with
\begin{equation}
\label{tail-exp}
\begin{array}{lll}
b=0 & \quad\mbox{inconsistent} & \quad\left(\mbox{GT};\:\: \nu=0\right) \\
b=\nu &\quad \beta =d \beta_\nu / \nu \varpi_2 & \quad\left(\mbox{GT};\:\:  \nu>0    \right) \\
b= \textstyle{\frac{1}{2}} \left( \nu +2\right)& \quad  \beta = \frac{2}{\nu+2} \sqrt{\frac{2d
\beta_\nu}{\varpi_2}} &\quad \left(\mbox{WN};\:\:  \nu > 0\ \right),
\end{array}
\end{equation}
where the scaling functions and high energy tails of GT-driven and
freely cooling systems are equivalent.

We conclude this subsection with some comments.\\
\noindent{\em driven systems:}\\
For $0<b<2$ both GT- and WN-driving lead to consistent asymptotic
solutions of the scaling equations for $\nu>0$ with overpopulated high
energy tails of stretched Gaussian type. We also note, the larger the
interactions (larger $\nu$-values) at large impact energies, the
smaller the overpopulation of the high energy tails. For $\nu=2$,
corresponding to the inelastic version of the Very Hard Particle model
(see end of Section 3.2) we obtain Gaussian behavior $(b=2)$ for both
types of thermostats, and there are no longer over-populated tails. In
the case of stretched Gaussian tails {\em all} moments $\int
d{\vec{c}} c^n f\left(c\right) <\infty$. This would not be the case
for power law tails.

In the case of white noise driving, the above results with $b=
1+\textstyle{\frac{1}{2}}\nu$, include inelastic hard spheres
$\left(\nu=1\right)$, as well as Maxwell models ($\nu=0$), and the
results coincide with the detailed predictions for $\nu=1$
\cite{TvN+ME-granmat} and $\nu=0$ \cite{ME+RB-rapid}. For GT-driven or
freely cooling IRS-models with $\nu >0$ the tail distributions have a
stretching exponent $b=\nu$. The property $\beta b \to 1/\gamma_0$ as
$\nu \downarrow 0$ would suggest power law behavior at the stability
threshold, $\nu=0$ (Maxwell models), with an exponent $a = 1/\gamma_0$
(see however next item). The result \eqref{tail-exp} also includes the
exponential decay, $f(c)\sim\exp[-\beta c]$ for inelastic hard
spheres, found in \cite{TvN+ME-granmat}.

\vspace{3mm} \noindent{\em Stability thresholds:}
\\ In the previous comments we have speculated on power law tails,
in agreement with the exact solution \cite{Rome1} for the
one-dimensional freely cooling Maxwell models with
$f\left(c\right)\sim c^{-4}$ for $c\gg 1$. As will be shown in section
5, the scaling form for $d$-dimensional Maxwell models in free cooling
does indeed have an algebraic tail $f\left(c\right)\sim 1/c^{d+a}$,
{\it but} with an exponent $a(\alpha) \neq 1/\gamma_0$. However,
neither the existence of algebraic tails, nor the possible value of
the power law exponents can be obtained legitimately from the results
\eqref{tail-exp}. The reason is that the result is not consistent with
the {\em a priori } assumption $I_{\mbox{gain}}\ll I_{\mbox{loss}} $
at $c\gg 1$. In fact, gain and loss terms are of the same order of
magnitude for $\nu=0$.

\vspace{3mm}\noindent{\em Maxwell models as approximations to hard spheres:}\\
As discussed in item 2 of Section 3.5, inelastic Maxwell models have
also been introduced as a sensibly looking  mathematical
simplification of the Boltzmann equation for inelastic hard spheres.
The results of the analysis in  the previous section show that the
effect of this simplification on the shape of the tail may be very
drastic. In the free cooling case the hard sphere tail is exponential,
$f\left(c\right) \sim \exp\left[-\beta c\right]$, whereas in the
Maxwell model it is a power law tail, $f\left(c\right) \sim
1/c^{d+a}$. A smaller difference exists in the WN-driven case, where
the hard sphere tail is $f\left(c\right) \sim \exp\left[-\beta
c^{3/2}\right]$, and the Maxwell tail is $f\left(c\right) \sim
\exp\left[-\beta c\right]$.

\vspace{3mm}\noindent{\em MC-simulations at high and low inelasticity:}\\
The results \eqref{tail-exp} predict more than the stretching exponent
$b$. In fact the coefficient $\beta$ for $\nu >0$ in $f\left(c\right)
\sim\exp\left[-\beta c^b\right]$ is given in terms of $\varpi_2$ in
\eqref{NESS-cond}.  For these $\nu-$values we have described a
perturbative calculation for $\varpi_2$, which converges rapidly for
$\alpha\to 1$, but gives poor results for $\alpha <0.6 $
\cite{TvN+ME-granmat,MS00}. Nevertheless, it is possible to test the
predictions of \eqref{tail-exp} for the GT- and the WN-thermostats,
including the coefficient $\beta$ for {\em all} values of the
restitution coefficient $\alpha$. This can be done by measuring the
mean square velocity $v_0^2\left(\infty\right)$ in MC simulations on
driven systems in a NESS for a given $\gamma$ or $D$. From these data
$\varpi_2$ can be calculated using \eqref{NESS-cond}.

\section{Inelastic Maxwell models}

\subsection{Fourier transform method}
As discussed in Section 4.3, the behavior of the high energy tails of
the scaling form in the IRS-models is only controlled by the loss term
in the Boltzmann collision term, and generates the stretched Gaussian
tails.   However, in the borderline case ($\nu=0)$ the tail behavior
is determined by an interplay between gain and loss term, which leads
to algebraic tails. This makes the analysis more complicated. In this
article we only discuss the borderline case in {\em free cooling},
formed by the Maxwell models, and we refer to \cite{ETB} for a more
comprehensive discussion.

In this section we want to demonstrate that the Boltzmann equation
\eqref{BE-nu} for inelastic Maxwell models ($\nu=0$),
\begin{equation}\label{BE-max}
\partial_t F\left(v,t\right)=  -F\left(v,t\right) + \frac{1}{\alpha} \int_{\vec{n}}
\int d{\vec{w}}  F\left(v^{**},t\right) F\left(w^{**},t\right)  ,
\end{equation}
has a scaling solution with a power law tail. To do so we first
consider the Fourier transform of the distribution function,
$\Phi\left(k,t\right) = \left<\exp\left[-i\vec{k} \cdot
{\vec{v}}\right]\right>$, which is the characteristic function or
generating function of the velocity moments. Because
$F\left(v,t\right)$ is isotropic, $\Phi\left(k,t\right)$ is isotropic
as well.

As an auxiliary step we first Fourier transform the gain term in
\eqref{BE-max}, i.e.
\begin{multline}
\label{fourier-gain}
\int d\vec{v} \exp\left[-i \vec{k}
\cdot \vec{v}\right] \:I_{\mbox{gain}}\left(v|F\right)=\\
\int_{\vec{n}} \int d\vec{v} d{\vec{w}} \exp\left[-i \vec{k} \cdot
\vec{v}^{*}\right] F\left(v,t\right) F\left(w,t\right)   =
\int_{\vec{n}} \Phi\left(k\eta_+,t\right)\Phi\left(k \eta_-,t\right).
\end{multline}
The transformation needed to obtain the first equality is the same as
in \eqref{m2-gen}. Then we use \eqref{v*} to write the exponent as
$\vec{k}\cdot \vec{v}^*_1 = \vec{k}_- \cdot \vec{v}_1 + \vec{k}_+
\cdot \vec{v}_2$, where
\begin{equation}
\label{eta-pm}
\begin{split}
\vec{k}_{+} &=  p \vec{k} \cdot \vec{n} \vec{n}  \qquad
\left| \vec{k}_+\right| =k p
\left|\left(\hat{\vec{k}}\cdot \vec{n}\right)\right| =
k \eta_+\left(\vec{n}\right) \\
\vec{k}_{-}&  =\vec{k} - \vec{k}_{+} \qquad  \left|\vec{k}_-\right| =
k\sqrt{1-z\left(\hat{\vec{k}}
 \cdot \vec{n}\right)^2} = k\eta_-\left(\vec{n}\right),
\end{split}
\end{equation}
with $p=1-q=\textstyle{\frac{1}{2}} \left(1+\alpha\right)$ and
$z=1-q^2$.  The Fourier transform of \eqref{BE-max} then becomes,
\begin{equation} \label{BE-fourier}
\partial_t \Phi\left(k,t\right)= - \Phi\left(k,t\right) +
\int_{\vec{n}} \Phi\left(k\eta_+\left(\vec{n}\right),t\right)
\Phi\left(k\eta_{-}\left(\vec{n}\right),t\right),
\end{equation}
where $\Phi\left(0,t\right)=1$ because of \eqref{low-m}. In
one-dimension $\eta_+\left(\vec{n}\right) =p $ and
$\eta_-\left(\vec{n}\right) =q $, and $\int_{\vec{n}} $ can be
replaced by unity. Moreover this equation simplifies to
\cite{BN-PK-00},
\begin{equation} \label{BE-1D-fourier}
\partial_t \Phi\left(k,t\right)= \Phi\left(p k,t\right)
\Phi\left(q k,t\right)-\Phi\left(k,t\right).
\end{equation}
Because $F\left(v,t\right)$ is isotropic, only the even moments are
non-vanishing, and the moment expansion  of the characteristic
function  takes the form,
\begin{equation} \label{charact}
\Phi\left(k,t\right)= {\sum_{n}}^\prime \frac{\left(-ik\right)^n}{n!}
\left<\left(\hat{\vec{k}}\cdot\vec{v}\right)^n\right>
={\sum_{n}}^\prime \left(-ik\right)^n m_n\left(t\right) ,
\end{equation}
where the prime indicates that $n=even$, and the moment
$m_n\left(t\right)$ is
defined as,
\begin{equation} \label{mom-n}
m_n\left(t\right) = \beta_n \left<v^n\right> /n!,
\end{equation}
where $\beta_n = \int_{\vec{n}} \left(\hat{\vec{k}}
\cdot\hat{\vec{v}}\right)^n $  is given in \eqref{beta-nu}. Moreover,
the normalizations \eqref{low-m} give $m_0\left(t\right)=1$ and $
m_2\left(t\right)= \textstyle{\frac{1}{2}} \beta_2 \left<v^2\right>
=\textstyle{\frac{1}{4}} v_0^2$.

Maxwell models have the unusual property that the system of moment
equations for $m_n\left(t\right)$ is closed, and can be solved {\em
sequentially}.  The reason is that the rate $\dot{m}_n$ is only a
function of lower moments $m_s\left(t\right)$ with $s \leq n$.

The moment equations are readily obtained by inserting the expansion
\eqref{charact} in the Fourier transformed Boltzmann equation
\eqref{BE-fourier}, and equating the coefficients of equal powers of
$k$. The result is,
\begin{equation} \label{mom-eq}
\dot{m}_n +\lambda_n m_n = \sum_{l=2}^{n-2} h\left(l,n-l\right)\:
m_l\: m_{n-l} \quad \left(n>2\right),
\end{equation}
where all labels $\{n,l,s\}$ take {\em even} values only. Following
\cite{BN+PK-02} or Appendix A of \cite{ME+RB-fest} the functions can
be calculated for real positive values of $\ell$ and $s$ with the
result,
\begin{equation}
\label{eigenval}
\begin{split}
  h\left(l,s\right) =& \int_{\vec{n}} \eta^l_+
\left(\vec{n}\right)\eta^s_-\left(\vec{n}\right)=
p^l \beta_{l}\:\: {_2}F_1\left(-\frac{s}{2},\frac{l+1}{2};\frac{l+d}{2}\:|\:z \right)\\
\lambda_s=& 1-h\left(s,0\right)-h\left(0,s\right)= \int_{\vec{n}}
\left[1-\eta^s_+\left(\vec{n}\right)-\eta^s_-\left(\vec{n}\right)\right]\\
=& 1-p^s\beta_s-\:{_2}F_1
\left(-\frac{s}{2},\frac{1}{2};\frac{d}{2}\:|\:1-q^2\right).
\end{split}
\end{equation}
Here ${_2}F_1(\alpha,\beta;\gamma\:|\:z)$ is a hypergeometric
function, and $\lambda_2 =2pq/d =2\gamma_0$ is easily obtained from
the above results.

Next we consider the Fourier transform of the scaling relation
\eqref{f-scale}, yielding $\Phi \left(k,t\right) =
\phi\left(kv_0\left(t\right)\right)$ with $v_0\left(t\right)=
v_0\left(0\right) \exp\left(-\gamma_0 t\right)$ according to
\eqref{zeta-0}. Inserting $\Phi$ in \eqref{BE-fourier} gives the
integral equation for the scaling form $\phi\left(k\right)$, which
reads
\begin{equation} \label{phi-scale-eq}
-\gamma_0 k \frac{d}{dk} \phi\left(k\right) +\phi\left(k\right) =
\int_{\vec{n}}\phi\left(k\eta_+\right)\phi\left(k\eta_-\right).
\end{equation}
For $d=1$ it reduces  to,
\begin{equation} \label{phi-scale-eq-1d}
-pq k \frac{d}{dk} \phi\left(k\right) +\phi\left(k\right) =
\phi\left(pk\right)\phi\left(qk\right).
\end{equation}
The scaling form $\phi \left(k\right)$ is the generating function for
the moments of  $f\left(c\right)$, i.e.
\begin{equation}
\begin{split}
\label{phi-k-exp} \phi\left(k\right) = &{\sum_n}^\prime
\frac{\left(-ik\right)^n}{n!} \beta_n \left<c^n\right> \equiv
{\sum_n}^\prime \left(-ik\right)^n \mu_n \\
\simeq & 1 -\textstyle{\frac{1}{4}} k^2 + k^4 \mu_4 - k^6\mu_6
+\cdots,
\end{split}
\end{equation}
 where $n=even$, $\mu_0=1$ and $ \mu_2 =
\textstyle{\frac{1}{2}}
\beta_2 \left<c^2\right>= 1/4$
on account of the normalizations \eqref{norm-scale} and $\beta_2 =1/d$
(see \eqref{beta-nu}). By inserting \eqref{phi-k-exp} into
\eqref{phi-scale-eq} one obtains the recursion relation,
\begin{equation} \label{recurs-mu}
\mu_n =\frac{1}{\lambda_n - \textstyle{\frac{1}{2}}n \lambda_2}
\sum_{l=2}^{n-2} h\left(l,n-l\right) \mu_l\mu_{n-l} \qquad
\left(n>2\right)
\end{equation}
with initialization $\mu_2=1/4$ and all labels $n,l=even$. How these
moments behave as a function of $\alpha$ has been calculated in
\cite{ME+RB-fest} by numerically solving the recursion relation.

\subsection{Small-$k$ singularity of  the characteristic function}

The asymptotic analysis in Section 4.4 of the high energy tail
$f\left(c\right)\sim \exp\left[-\beta c^b\right]$ for Maxwell models
($\nu=0$) yields $b=0$, which suggests a power law tail
$f\left(c\right) \sim 1/c^{d+a}$ as the leading large$-c$ behavior in
$d-$dimensional systems.  If this is indeed the case, then the moments
$\mu_n$ of the scaling form $f\left(c\right)$ are convergent if $n<a$
and are divergent if $n> a$. As we are interested in physical
solutions which can be {\em normalized}, and have a {\em finite
energy}, a possible value of the power law exponent must obey $a>2$.

The characteristic function is in fact a very suitable tool for
investigating this problem. Suppose the moment $\mu_n$ with $n>a$
diverges, then the $n$-th derivative of the corresponding generating
function also diverges at $k=0$, i.e. ${\phi}\left(k\right)$ has a
singularity at $k=0$. Then a simple rescaling argument of the inverse
Fourier transform shows that $\phi\left(k\right)$ has a dominant
small-$k$ singularity of the form  ${\phi}\left(k\right) \sim
\left|k\right|^a$, where $a \neq even$. On the other hand, whenever
the exponent $b$ in $f\left(c\right)\sim \exp\left[-\beta c^b\right]$
is positive, -- as is the case for the inelastic Maxwell model {\it
driven by white noise},  where  $ b=1$ and $ \beta =\sqrt{2/pq}$ -- ,
then all moments are finite, and the characteristic function
$\phi\left(k\right) $ is regular at the origin, i.e. can be expanded
in powers of $k^2$.

We first illustrate our analysis for the one-dimensional case. As the
requirement of finite total energy imposes the lower bound $a>2$ on
the exponent, we make the ansatz, consistent with \eqref{phi-k-exp} that
the dominant small-$k$ singularity has the form,
\begin{equation}\label{ansatz}
\phi\left(k\right) = 1- \textstyle{\frac{1}{4}} k^2 + A\left|k\right|^a,
\end{equation}
insert this in \eqref{phi-scale-eq-1d}, and equate the coefficients of
equal powers of $k$. This yields,
\begin{equation} \label{transc-1d-eq}
\textstyle{\frac{1}{2}}a \lambda_2 = \lambda_a \quad {\rm or} \quad a
pq = 1-p^a-q^a.
\end{equation}
The equation has two roots, $a=2,3$, of which $a=3$ is the one larger
than 2. Here $A$ is left undetermined. Consequently the
one-dimensional scaling solution has a power law tail,
${f}\left(c\right) \sim 1/c^4$, in agreement with the exact solution
in \cite{Rome1}.

For general dimension we proceed in the same way as in the
one-dimensional case, insert the ansatz \eqref{ansatz} into
\eqref{phi-scale-eq}, and equate the coefficients of equal powers of
$k$. This yields for the coefficient of $k^2$ the identity $2\gamma_0
=\lambda_2$, and for the coefficient of $k^a$ the transcendental
equation,
\begin{equation} \label{trans-eq}
\textstyle{\frac{1}{2}}a \lambda_2   =\lambda_a =\int_{\vec{n}}
\left[1-\eta^a_+ -\eta^a_-\right].
\end{equation}
The equation above obviously has the solution $a=2$. We are
however interested in the solution with $a>2$. In the elastic
limit ($\alpha \to 1$) the solution is simple. There $ \gamma_0
\to 0$ and $a$ diverges. The contributions of $\eta_\pm^a$ on the
right hand side vanish because $\eta_\pm <1$, and the result is,
\begin{equation} \label{a-infty}
a \simeq \frac{1}{\gamma_0} = \frac{4d}{1-\alpha^2}.
\end{equation}
For general values of $\alpha$ one can conveniently use an integral
representation of $_2F_1$ to evaluate $\lambda_a$ and solve the
transcendental equation  \eqref{trans-eq} numerically. We illustrate
the solution method of \eqref{trans-eq} with the graphical
construction in Figure 2, where we look for intersections of the line
$y=\frac{1}{2} \lambda_2= \gamma_0 s$ with the curve $y=\lambda_s$ for
different values of $\alpha$.

\begin{figure}[htbp]
  $$\includegraphics[width=.65 \columnwidth]{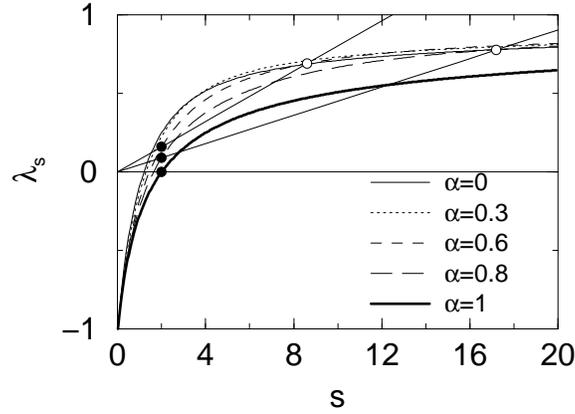}$$
\caption{Graphical solution of Eq.\protect{\eqref{trans-eq}} for
different values of the parameter $\alpha$. The eigenvalue
$\lambda_s$ is a concave function of $s$, plotted for different
values of the restitution coefficient $\alpha$ for the 2-D
inelastic Maxwell model. The line $y=s\gamma_0$ is plotted for
$\alpha=0.6,0.8$ and $\alpha =1$(top to bottom). The intersections
with $\lambda_s$ determine the points $s_0$ (filled circles) and
$s_1$ (open circles). Here $s_1=a$ determines the exponent of the
power law tail. For the elastic case ($\alpha=1, \gamma_0=0$,
energy conservation) there is only one intersection point.}
\end{figure}

The relevant properties  of $\lambda_s $ are: (i)  $\lim_{s \to 0}
\lambda_s =-1$; (ii) $\lambda_s$ is a concave function, monotonically
increasing with $s$, and {\em (iii)} all eigenvalues for positive {\em
integers} $n$ are positive (see Figure 2). As can be seen from the
graphical construction, the transcendental equation \eqref{trans-eq}
has two solutions, the trivial one $\left(s_0=2\right)$ and the
solution $s_1=a$ with $a>2$. The numerical solutions for $d=2,3$ are
shown in Figure 3 as a function of $\alpha$, and the
$\alpha$-dependence of the root $a\left(\alpha\right)$ can be
understood from the graphical construction. In the elastic limit as
$\alpha \uparrow 1$ the eigenvalue $\lambda_2\left(\alpha\right)\to 0$
because of energy conservation. In that limit the transcendental
equation \eqref{trans-eq} no longer has a solution with $a>2$, and
$a\left(\alpha\right)\to\infty$ according to \eqref{a-infty}, as it
should be. This is consistent with a Maxwellian tail distribution in
the elastic case. Krapivsky and Ben-Naim have in fact solved the
transcendental equation asymptotically for large $d$, which gives
qualitatively the same results as shown in Figure 3 for two and three
dimensions.

\begin{figure}[htbp]
 $$\includegraphics[angle=270,width=.65 \columnwidth]{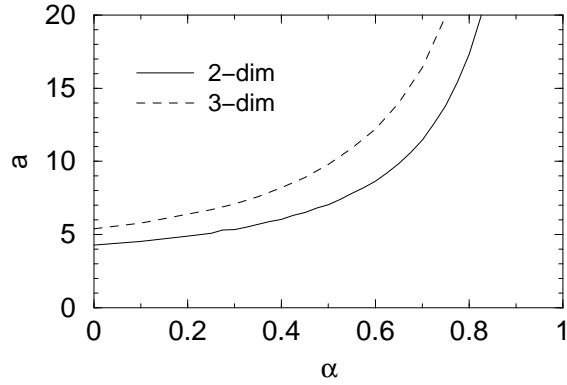}$$
\caption{Exponent $ a\left(\alpha\right)$, which is the non-trivial
root of \eqref{trans-eq}, shown as a function of the coefficient of
restitution $\alpha$, and denoted in the plot by $\alpha$. It
determines the high energy tail $1/c^{a+d}$ of the scaling solution
$f\left(c\right)$ of the 2-D and 3-D Maxwell models. }
\end{figure}

These results establish the existence of scaling solutions
$f\left(c\right) \sim 1/c^{d+a}$ with algebraic tails, where the
exponent $a$ is the solution of the transcendental equation
\eqref{trans-eq} with $a>2$. The exponent $a\left(\alpha\right)$
behaves as a function of $\alpha$ qualitatively the same as in the
simple inelastic BGK model of Section 2.

Using a somewhat different analysis Krapivsky and Ben-Naim
\cite{BN+PK-6-11} independently obtained the same results for the
algebraic tails in freely cooling Maxwell models.

\subsection{Beyond asymptotic analysis}
\noindent {\em Free cooling:}\\
As mentioned already in Section 3.3, Baldassarri et al. \cite{Rome1}
have obtained an exact solution $f\left(c\right)$ of the scaling
equation for the one-dimensional Maxwell model in free cooling, and
they demonstrated the importance of this scaling solution by means of
Monte Carlo simulations. In doing so they found  that
$F\left(v,t\right)$ for different classes of initial distributions can
be collapsed for {\em long } times $t$ on this exact  scaling solution
$f\left(c\right)$, when $v_0\left(t\right) F\left(v,t\right)$ is
plotted versus $c= v/v_0\left(t\right)$.

We briefly illustrate here how this solution is obtained from the
one-dimensional scaling equation \eqref{phi-scale-eq-1d}. One first
verifies that the following function,
\begin{equation} \label{exact-phi}
\phi\left(k\right) = \left(1+\vartheta \left|k\right|\right)
\exp\left(-\vartheta \left|k\right|\right),
\end{equation}
with arbitrary positive $\vartheta$ is a solution, that can be
Fourier inverted. It gives the scaling form,
\begin{equation} \label{exact-f}
f\left(c\right) = \frac{2}{\pi \vartheta} \frac{1}{\left(1+
c^2/\vartheta^2\right)^2}.
\end{equation}
To determine the scaling form that satisfies the normalizations
\eqref{norm-scale}, we expand \eqref{exact-phi} in powers of
$\left|k\right|$ to obtain,
\begin{equation} \label{k-exp-exact}
\phi\left(k\right) = 1 -\textstyle{\frac{1}{2}} \vartheta^2 k^2 +
\textstyle{\frac{1}{3}} \vartheta^3 \left|k\right|^3
+\cdots.
\end{equation}
Comparison of this result with \eqref{phi-k-exp} shows that $\vartheta
=1/\sqrt{2}$. Moreover, it confirms the ansatz \eqref{ansatz} used to
find solutions with small-$k$ singularities. Comparison also shows the
value of the coefficient $A=1/\left[6\sqrt{2}\right]$ in
\eqref{ansatz}, and the high energy tail is $f\left(c\right) \sim
{\cal A}/c$ with ${\cal A} = 1/\left[\pi \sqrt{2}\right]$. The
coefficients $A$ and ${\cal A}$ can not be determined within the
asymptotic method.

\vspace{3mm}\noindent {\em White noise driving:}\\ For the
one-dimensional Maxwell model driven by white noise the steady state
solution has also been found exactly with the help of the Fourier
transform method \cite{BN-PK-00}. The characteristic function
satisfies in that case a nonlinear finite difference equation, which
can be solved by iteration. However, the analytic structure is rather
complex, and makes it difficult to extract analytic information from
that solution.

The observation that $f\left(c\right) \sim \exp\left[-\beta
\left|c\right|\right]$ has an exponential high energy tail, was
probably first made in numerical work of van der Hart and Nienhuis
\cite{Nienhuis}. A more detailed analytic prediction about the
asymptotic tail was given in \cite{ME+RB-rapid}, where it was shown
that $\beta = \sqrt{8/\left(1-\alpha^2\right)}$ for all Maxwell
models, independent of the dimensionality $\left(d=1,2,\cdots\right)$,
at least with the normalization, $\left<c^2\right>=
\textstyle{\frac{1}{2}} d$, used in this article. This result follows
directly from \eqref{tail-exp} and the value $\varpi_2 = d \gamma_0
=\textstyle{\frac{1}{4}} \left(1-\alpha^2\right)$, given in item 3
below \eqref{NESS-cond}. Furthermore, additional numerical and
analytical work was also published by Marconi and Puglisi
\cite{Rome3}, and  by Antal et al. \cite{Droz}. Only recently more
detailed analytic results have been extracted from the rather complex
structure of the exact solution \cite{BN-PK-Springer,AS+ME-condmat}.
\section{Conclusions and perspectives}

We have studied asymptotic properties of scaling or similarity
solutions, $ F\left(v,t\right) = \left(v_0\left(t\right)\right)^{-d}
f\left(v/v_0\left(t\right)\right)$, of the nonlinear Boltzmann
equation in spatially homogeneous systems composed of particles with
{\it inelastic} interactions for large times and large velocities. The
large $t-$ and $v-$scales are relevant because on such scales the
universal features of the solutions survive, while details of the
initial distributions, of interaction strength, and degree of
inelasticity, are mostly lost. The behavior of these scaling states,
which describe nonequilibrium steady states (NESS), is less universal
than the state of thermal equilibrium, because the form of the NESS
distribution $f(c)$ depends on the way of driving the dissipative
systems. Scaling solutions are very well suited to expose the
universal features of the velocity distribution functions, because the
velocities, $c= v/v_0\left(t\right)$, are measured in units of the
r.m.s. velocity or instantaneous width $v_0\left(t\right)$ of the
distribution.

The real importance of the scaling solutions is that the actual
solutions $F(v,t)$ for large classes of initial distributions $F(v,0)$
(essentially initial data without over-populated tails -see
\cite{ME+RB-fest}) rapidly approach these scaling solutions in the
sense that after a short transient time the data $v_0^d\left(t\right)
F\left(v,t\right) $ can be collapsed on a single scaling form
$f\left(c\right)$, as first observed by Baldassarri et al. in MC
simulations of the nonlinear Boltzmann equation for a one-dimensional
inelastic Maxwell model. This rapid approach to a universal scaling
function applies both to systems driven by a Gaussian or by a white
noise thermostat, as well as for freely cooling systems, which show
scaling behavior, identical to systems driven by a Gaussian
thermostat.

Originally this scenario had the status of a conjecture for systems of
inelastic hard spheres ($\nu=1$) and inelastic Maxwell
models($\nu=0$), as formulated in \cite{ME+RB-fest} where also some
analytical evidence for the approach to a scaling form has been
presented. For Maxwell models the conjecture has been rigorously
proven in the mean time by Bobylev et al. \cite{AB+CC-proof}. The
analysis in Section 4.1 also suggests what the basic criterion us for
the approach of IRS-models to the scaling form. If the energy balance
equation \eqref{vo-driven} has a stable/attractive fixed point
solution, $v_0(\infty)$, then the distribution function $F(v,t)$
approaches -- in the sense detailed in the paragraphs above -- a
stable NESS described by the scaling solution.

The analysis in Section 4.1 of the energy balance equation also
suggests that cases of marginal stability are candidates for power law
tails, where the freely evolving inelastic Maxwell model is a well
known example. A further candidate system for power law tails is an
IRS-model with $\nu=-2$, driven by white noise. These properties, and
the relation between marginal stability and power law tails is further
explored in \cite{ETB}, where in addition the analytic predictions of
the present paper are confirmed by MC simulations.

In this article we have focussed on the properties of the scaling
solution, and in particular on its high energy tail. Here we have
introduced the Boltzmann equation for new classes of inelastic
interactions, named Inelastic Repulsive Scatterers or IRS-models,
corresponding to pseudo-repulsive power law potentials, with collision
probabilities proportional to $\textsc{a} \sim \left|\vec{g} \cdot
\vec{n}\right|^\nu$, covering hard scatterers like inelastic hard
spheres ($\nu =1$) and soft scatterers like pseudo-Maxwell molecules
($\nu=0$).  The energy loss in an inelastic interaction is
proportional to the inelasticity, $\gamma_0 \sim
\left(1-\alpha^2\right)$, where $\alpha$ is the coefficient of
restitution. We have studied two typical cases: freely evolving
systems in homogeneous cooling states, without energy supply, and
systems with energy supply, driven by Gaussian thermostats with
negative friction or driven by Gaussian white noise.

The homogeneous cooling laws in these systems are described by the
granular temperature $T=v_0^2$, with a long time decay as $t^{-2/\nu}$
or $\exp\left[-2\gamma_\nu \tau\right]$, where $t$ is the external
(laboratory) time, and $\tau$ is the internal time or collision
counter, and $\gamma_\nu$ is proportional to the inelasticity
$\gamma_0$.

An interesting feature of all IRS-models is that their scaling forms
generically have a stretched Gaussian high energy tail,
$f\left(c\right) \sim \exp[- \beta c^b]$. In {\em freely cooling}
IRS-models with $\nu>0$ the stretching exponent $b=\nu$ satisfies
$0<b=\nu \leq 2 $. Among the IRS-models at the stability threshold
only the freely cooling Maxwell model has been analyzed in the present
paper, and it does yield power law tails, $f\left(c\right) \sim
1/c^{d+a}$, and the exponent $a$ has been calculated from a
transcendental equation. In the IRS-models with $\nu>0$, driven by
white noise, there exist again stretched Gaussian tails with
$b=\textstyle{\frac{1}{2}} \left(\nu +2\right)$. All known results for
higher dimensional inelastic hard spheres ($\nu=0$) and inelastic
Maxwell models ($\nu=1$) are recovered in the present analysis.

In IRS-models  with positive $\nu$ the  tails are always stretched
Gaussians, and they are determined only by the loss term in the
Boltzmann equation. In the freely cooling Maxwell model, and more
generally at a stability threshold  -- which are the cases leading to
power law tails \cite{ETB} -- the loss and gain term  in the nonlinear
Boltzmann equation are of comparable size, and partially balancing
each other.

We also have analyzed  in Section 2 an inelastic BGK model, introduced
by Brey et al \cite{BMD}, and generalized it to an energy dependent
collision frequency, and we added energy source terms to this kinetic
equation as well. In all cases the scaling solution can be calculated
exactly. The freely cooling model shows an algebraic tail,
$f\left(c\right) \sim 1/c^{d+a}$ with an exponent $a \sim 1/\gamma_0
\sim 2/\left(1-\alpha^2\right)$. For white noise driving one finds
asymptotically $f \left(c\right) \sim \exp\left[-\beta c\right]$ with
$\beta \sim \left(1-\alpha^2\right)^{-1/2}$.

It is interesting to observe that the overpopulated high energy tails
in the inelastic BGK-models, both for free cooling as well as for
white noise driving, are essentially the same as for the more
complicated Maxwell models. However, the present extension to an
energy dependent collision frequency  does not capture the generic
features of the less-singular stretched Gaussian tails for the
IRS-models with $\nu>0$ , as predicted by the nonlinear Boltzmann
equation for these models.

\section*{Acknowledgements}
The authors want to thank J.W. Dufty, A. Santos, E. Trizac, A. Barrat
and K. Shundyak for helpful discussions and correspondence. This work
is supported by DGES (Spain), Grant No BFM-2001-0291.

\end{document}